\documentclass[english]{article}
\usepackage{amsfonts}
\usepackage{amsmath}
\usepackage{amssymb}
\usepackage{amsthm}
\usepackage{verbatim}
\usepackage{float}
\usepackage{subfig}
\usepackage{multirow}
\usepackage[colorlinks,citecolor=blue,urlcolor=blue,linkcolor=red]{hyperref}
\usepackage{fullpage}
\usepackage[affil-it]{authblk}
\usepackage{siunitx}
\usepackage{textcomp}
 \usepackage{tabularx}
\usepackage{booktabs}
\usepackage{caption}
\usepackage{graphicx}
\usepackage[normalem]{ulem}
\useunder{ine}{}{}
\usepackage{lscape}

\renewcommand{\phi}{\varphi}
\renewcommand{\epsilon}{\varepsilon}

\begin{document}
\title{Towards Label-Free 3D Segmentation of Optical Coherence Tomography Images of the Optic Nerve Head Using Deep Learning}

\author[1]{Sripad Krishna Devalla}
\author[1,2]{Tan Hung Pham}
\author[1]{Satish Kumar Panda}
\author[1]{Liang Zhang}
\author[1]{Giridhar Subramanian}
\author[1]{Anirudh Swaminathan}
\author[1]{Chin Zhi Yun}
\author[3]{Mohan Rajan}
\author[3]{Sujatha Mohan}
\author[4]{Ramaswami Krishnadas}
\author[4]{Vijayalakshmi Senthil}
\author[5]{John Mark S. de Leon}
\author[1,2]{Tin A. Tun}
\author[2,8]{Ching-Yu Cheng}
\author[2,6,9,10,11]{Leopold Schmetterer}
\author[2,7]{Shamira Perera}
\author[2,7]{Tin Aung}
\author[12$\star$]{Alexandre H. Thi\'{e}ry}
\author[1,2 $\alpha$]{Micha\"el J. A. Girard}

\affil[1]{Ophthalmic Engineering and Innovation Laboratory, Department of Biomedical Engineering, Faculty of Engineering, National University of Singapore, Singapore.}
\affil[2]{Singapore Eye Research Institute, Singapore National Eye Centre, Singapore.}
\affil[3]{Rajan Eye Care Hospital, Chennai, India.}
\affil[4]{Glaucoma Services, Aravind Eye Care Systems, Madurai, India.}
\affil[5]{Department of Health Eye Center, East Avenue Medical Center, Quezon City, Philippines.}
\affil[6]{Nanyang Technological University, Singapore.}
\affil[7]{Duke-NUS Graduate Medical School.}
\affil[8]{Ophthalmology and Visual Sciences Academic Clinical Program (Eye ACP), Duke-NUS Medical School, Singapore}
\affil[9]{Department of Clinical Pharmacology, Medical University of Vienna, Austria.}
\affil[10]{Center for Medical Physics and Biomedical Engineering, Medical University of Vienna, Austria.}
\affil[11]{Institute of Clinical and Molecular Ophthalmology, Basel, Switzerland}
\affil[12]{Department of Statistics and Applied Probability, National University of Singapore, Singapore.}

\bigskip

\affil[$\star$ $\alpha$]{Both contributed equally and both are corresponding authors.}

\affil[$\alpha$]{mgirard (at) invivobiomechanics.com}
\affil[$\star$]{a.h.thiery (at) nus.edu.sg}

\maketitle


%
%
\begin{abstract}
\noindent

\textbf{}
Since the introduction of optical coherence tomography (OCT), it has been possible to study the complex 3D morphological changes of the optic nerve head (ONH) tissues that occur along with the progression of glaucoma. Although several deep learning (DL) techniques have been recently proposed for the automated extraction (segmentation) and quantification of these morphological changes, the device-specific nature and the difficulty in preparing manual segmentations (training data) limit their clinical adoption. With several new manufacturers and next-generation OCT devices entering the market, the complexity in deploying DL algorithms clinically is only increasing. To address this, we propose a DL-based 3D segmentation framework that is easily translatable across OCT devices in a label-free manner (i.e. without the need to manually re-segment data for each device). Specifically, we developed 2 sets of DL networks. The first (referred to as the ‘enhancer’) was able to enhance OCT image quality from 3 OCT devices, and harmonized image-characteristics across these devices. The second performed 3D segmentation of 6 important ONH tissue layers. We found that the use of the ‘enhancer’ was critical for our segmentation network to achieve device independency. In other words, our 3D segmentation network trained on any of 3 devices successfully segmented ONH tissue layers from the other two devices with high performance (Dice coefficients $>$ 0.92). With such an approach, we could automatically segment images from new OCT devices without ever needing manual segmentation data from such devices. 

\end{abstract}

\section{Introduction}
\par The complex 3D structural changes of the optic nerve head (ONH) tissues that manifest with the progression of glaucoma has been extensively studied and better understood owing to the advancements in optical coherence tomography (OCT) imaging \cite{RN1}. These include changes such as the thinning of the retinal nerve fiber layer (RNFL) \cite{RN2,RN3}, changes in the choroidal thickness \cite{RN4}, minimum rim width \cite{RN5}, and lamina curvature and depth \cite{RN6,RN7}. The automated segmentation and analysis of these parameters in 3D from OCT volumes could improve the current clinical management of glaucoma.\\
\par Robustly segmenting OCT volumes remains extremely challenging. While commercial OCTs have in-built proprietary segmentation software, they can segment some, but not all the ONH tissues \cite{RN97,RN98,RN96}. To address this, several research groups have developed an overwhelming number of traditional image processing based 2D \cite{RN8,RN9,RN10,RN11,RN12,RN13} and 3D \cite{RN14,RN15,RN16,RN17,RN18,RN19} segmentation tools, they are generally tissue-specific \cite{RN8,RN10,RN11,RN13,RN14,RN19}, computationally expensive \cite{RN18,RN20}, require manual input \cite{RN15,RN17}, and are often prone to errors in scans with pathology \cite{RN18,RN21,RN22}.\\
\par Recent deep learning (DL) based systems have however exploited a combination of low- (i.e. edge-information, contrast and intensity profile) and high-level features  (i.e. speckle pattern, texture, noise) from OCT volumes to identify different tissues, yielding human-level \cite{RN23,RN24,RN25,RN26,RN27,RN28,RN29,RN30} and pathology invariant \cite{RN23,RN24,RN29} segmentations. Yet, given the variability in image characteristics (e.g. contrast or speckle noise) across devices as a result of proprietary processing software \cite{RN31}, a DL system designed for one device cannot be directly translated to others \cite{RN32}. Since it is common for clinics to own different OCT devices, and for patients to be imaged by different OCT devices during their care, the device-specific nature of these DL algorithms considerably limit their clinical adoption.\\
\par While there currently exists only a few major commercial manufacturers of spectral-domain OCT (SD-OCT) such as Carl Zeiss Meditec (Dublin, CA, USA), Heidelberg Engineering (Heidelberg, Germany), Optovue Inc. (Fremont, CA, USA), Nidek (Aichi, Japan), Optopol Technology (Zawiercie, Poland), Canon Inc. (Tokyo, Japan), Lecia Microsystems (Wetzlar, Germany), etc., several others have already started to or will soon be releasing the next-generation OCT devices. This further increases the complexity in deploying DL algorithms clinically. Given that reliable segmentations \cite{RN31} are an important step towards diagnosing glaucoma accurately, there is a need for a single DL segmentation framework that is not only translatable across devices, but also versatile to accept data from next-generation OCT devices.\\
\par In this study, we developed a DL-based 3D segmentation framework that is easily translatable across OCT devices in a label-free manner (without the need to manually re-segment data for each device). To achieve this, we first designed an ‘enhancer’: a DL network that can improve the quality of OCT B-scans and harmonize image characteristics across OCT devices. Because of such pre-processing, we demonstrate that a segmentation framework trained on one device can be used to segment volumes from other unseen devices.\\

\section{Methods}
\subsection{Overview}
\par The proposed study consisted of two parts: \textbf{(1)} image enhancement, and \textbf{(2)} 3D segmentation.\\
\par We first designed and validated a DL based image enhancement network to simultaneously de-noise (reduce speckle noise), compensate (improve tissue visibility and eliminate artefacts) \cite{RN34}, contrast enhance (better differentiate tissue boundaries) \cite{RN34}, and histogram equalize (reduce intensity inhomogeneity) OCT B-scans from three commercially available SD-OCT devices (Spectralis, Cirrus, RTVue). The network was trained and tested with images from all three devices.\\
\par A 3D DL-based segmentation framework was then designed and validated to isolate six ONH tissues from OCT volumes. The framework was trained and tested separately on OCT volumes from each of the three devices with and without image enhancement.


\subsection{Patient Recruitment}
\par A total of 450 patients were recruited from four centers: the Singapore National Eye Center (Singapore), Rajan Eye Care Hospital (Chennai, India), Aravind Eye Hospital (Madurai, India), and the East Avenue Medical Center (Quezon City , Philippines) \textbf{Table \ref{tab:table_1}}. All subjects gave written informed consent. The study adhered to the tenets of the Declaration of Helsinki and was approved by the institutional review board of the respective hospitals. The cohort comprised of 225 healthy and 225 glaucoma subjects. The inclusion criteria for healthy subjects were: an intraocular pressure (IOP) less than 21 mmHg, healthy optic discs with a vertical cup-disc ratio (VCDR) less than or equal to 0.5, and normal visual fields tests. Glaucoma was diagnosed with the presence of glaucomatous optic neuropathy (GON), VCDR $>$ 0.7 and/or neuroretinal rim narrowing with repeatable glaucomatous visual field defects. We excluded subjects with corneal abnormalities that could preclude the quality of the scans.\\

\subsection{Optical Coherence Tomography Imaging}

\par All 450 subjects were seated and imaged using spectral-domain OCT under dark room conditions in the respective hospitals. 150 subjects (75 healthy + 75 glaucoma) had one of their ONHs imaged using Spectralis (Heidelberg Engineering, Heidelberg, Germany), 150 (75 healthy + 75 glaucoma) using Cirrus (model: HD 5000, Carl Zeiss Meditec, Dublin, CA, USA), and another 150 (75 healthy + 75 glaucoma) using RTVue (Optovue Inc., Fermont, CA, USA). For glaucoma subjects, the eye with GON was imaged, and if both eyes met the inclusion criteria, one eye was randomly selected. For healthy controls, the right ONH was imaged. The scanning specifications for each device can be found in \textbf{Table \ref{tab:table_1}}.\\

\par From the dataset of 450 volumes, 390 (130 from each device) were used for training and testing the image enhancement network, while the remaining 60 (20 from each device) were used for training and testing the 3D segmentation framework.

\begin{table}[ht]
\centering
\resizebox{\textwidth}{!}{%
\begin{tabular}{cllll}
\hline
\multirow{2}{*}{\textbf{Device}} &
  \multicolumn{1}{c}{\multirow{2}{*}{\textbf{Hospital}}} &
  \multicolumn{2}{c}{\textbf{No of subjects}} &
  \multicolumn{1}{c}{\multirow{2}{*}{\textbf{Scanning Specifications}}} \\ \cline{3-4}
 &
  \multicolumn{1}{c}{} &
  \multicolumn{1}{c}{\textbf{Normal}} &
  \multicolumn{1}{c}{\textbf{Glaucoma}} &
  \multicolumn{1}{c}{} \\ \hline
\multicolumn{1}{l}{\textbf{Spectralis}} &
  \begin{tabular}[c]{@{}l@{}}Singapore National Eye Center\\ \\ Aravind Eye Hospital\end{tabular} &
  \begin{tabular}[c]{@{}l@{}}57\\ \\ 18\end{tabular} &
  \begin{tabular}[c]{@{}l@{}}11\\ \\ 64\end{tabular} &
  \begin{tabular}[c]{@{}l@{}}97 horizontal B-scans\\ 32 $\mu$m distance between B-scans\\ area of 15 $^{\circ}$ x 10 $^{\circ}$\\ centered on the ONH\\ 20x signal averaging.\end{tabular} \\ \hline
\textbf{Cirrus} &
  Rajan Eye Care Hospital &
  75 &
  75 &
  \begin{tabular}[c]{@{}l@{}}200 horizontal B-scans\\ 30 $\mu$m, 200 A-scans per B-scans\\ area of 6mm x 6mm centered on the ONH\end{tabular} \\ \hline
\textbf{RTVue} &
  East Avenue Medical Center &
  75 &
  75 &
  \begin{tabular}[c]{@{}l@{}}101 horizontal B-scans\\ 40 $\mu$m distance between B-scans\\ 101 A-scans per B-scan\\ area of 20 $^{\circ}$ x 20 $^{\circ}$ centered on the ONH\end{tabular} \\ \hline
\end{tabular}%
}
\caption{A summary of patient populations and scanning specifications for each OCT device.}
\label{tab:table_1}
\end{table}

\subsection{Image Enhancement}
The enhancer network was trained to reproduce simple mathematical operations including spatial averaging, compensation, contrast enhancement, and histogram equalization. When using images from a single device, the use of a DL network to perform such operations would be seen as unnecessary, as one could readily use the mathematical operators instead. However, when mixing images from multiple devices, besides performing such enhancement operations, the network also reduces the differences in the image characteristics across the devices, resulting in images that are ‘harmonized’ (i.e. less device specific) – a necessary step to perform robust device-independent 3D segmentation.\\

\begin{figure}[!htp]
    \centering
    \includegraphics[scale=0.35]{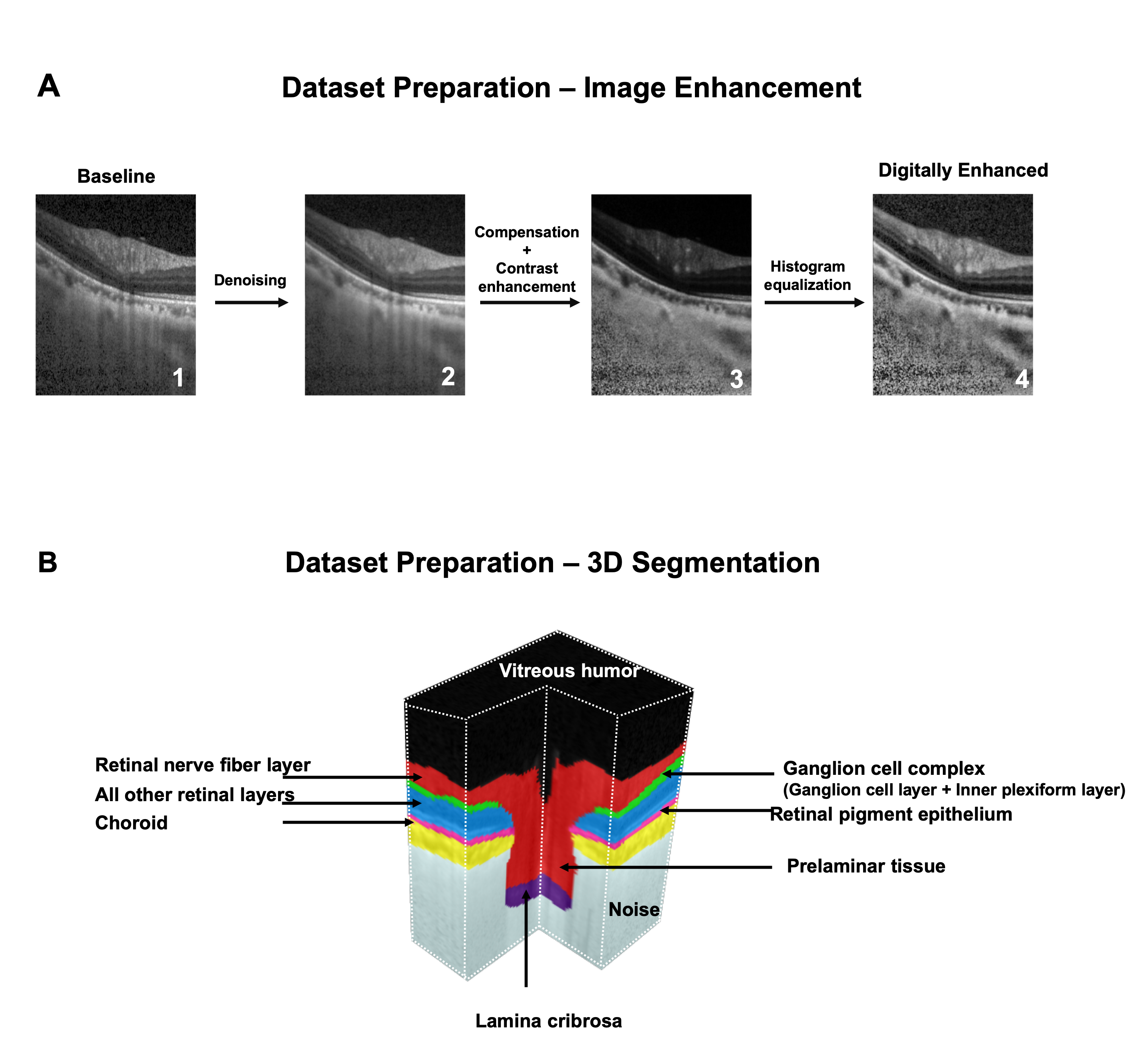}
    \caption{The dataset preparation for the image enhancement network is shown in \textbf{(A)}. Each B-scan (A [1]) was digitally enhanced (4) by performing spatial averaging (each pixel value was replaced by the mean of its 8 lateral neighbors; A [2]), compensation and contrast enhancement (contrast exponent = 2; A[3]), and histogram equalization (contrast limited adaptive histogram equalization [CLAHE], clip limit = 2; A[4]). For training the 3D segmentation framework \textbf{(B)}, the following tissues were manually segmented from OCT volumes: (1) the RNFL and prelamina (in red), (2) the ganglion cell complex (GCC; ganglion cell layer + inner plexiform layer; in cyan), (3) all other retinal layers (in blue); (4) the retinal pigment epithelium (RPE; in pink); (5) the choroid (in yellow); and (6) the lamina cribrosa (LC; in indigo). Noise (in grey) and vitreous humor (in black) were also isolated. }
    \label{fig:figure_1}
\end{figure}


\subsection{Image Enhancement – Dataset Preparation}
\par The 390 volumes were first resized (in pixels) to 448 (height) x 352 (width) x 96 (number of B-scans), and a total of 37,440 baseline B-scans (12,480 per device) were obtained. Each B-scan (\textbf{Figure \ref{fig:figure_1} A[1]}) was then digitally enhanced (\textbf{Figure \ref{fig:figure_1} A[4]}) by performing spatial averaging (each pixel value was replaced by the mean of its 8 lateral neighbors; \textbf{Figure \ref{fig:figure_1} A[2]}) \cite{RN35}, compensation and contrast enhancement (contrast exponent = 2; \textbf{Figure \ref{fig:figure_1} A[3]}) \cite{RN34}, and histogram equalization (contrast limited adaptive histogram equalization [CLAHE], clip limit = 2; \textbf{Figure \ref{fig:figure_1} A[4]}) \cite{RN36}.\\
\par The image enhancement network was then trained with 36,000 pairs (12,000 per device) of baseline and digitally-enhanced B-scans, respectively. Another 1,440 pairs were used for testing. B-scans from a same patient were not shared between training and testing.\\

\subsection{Image Enhancement – Network Description}
\par Briefly, as in our earlier DL based image enhancement study \cite{RN37}, the proposed enhancer exploited the inherent advantages of U-Net38 and its skip connections \cite{RN39}, residual learning \cite{RN40}, dilated convolutions \cite{RN41}, and multi-scale hierarchical feature extraction \cite{RN42}. We used the same network architecture, except that the output layer was now activated by the sigmoid activation function \cite{RN43} (originally tanh). The design, implementation, significance of each component, and data augmentation details can be referred to from our earlier study \cite{RN37}. The loss function was a weighted combination of both the root mean square error (RMSE) and a multi-scale perceptual loss \cite{RN44} function that was based on the VGG19 DL model \cite{RN45}.\\
\par Pixel-to-pixel loss functions (e.g., RMSE) compare only the low-level features (i.e., edge information) between the DL prediction and their corresponding ground-truth often leading to over-smoothened (blur) images \cite{RN44}, especially in image-to-image translation problems (e.g., de-noising). However, perceptual loss based functions exploit the high-level features (i.e., texture, abstract patterns) \cite{RN44,RN46,RN47,RN48} in these images to assess their differences, enabling the DL network to achieve human-like visual understanding \cite{RN49}. Thus, a weighted combination of both the loss functions allows the DL network to preserve the low- and high- level features in its predictions, limiting the effects of blurring.\\
\par To compute the perceptual loss, the output of the enhancer (referred to as 'DL-enhanced' B-scan) and its corresponding digitally-enhanced B-scan was separately passed through the VGG-19 \cite{RN45} DL model that was pre-trained on the ImageNet dataset \cite{RN50}. Feature maps at multiple scales (outputs from the 2nd, 4th, 6th, 10th, and 14th convolutional layers) were extracted, and the perceptual loss was computed as the mean RMSE (average of all scales) between the extracted features from the ‘DL-enhanced’ and its corresponding ‘digitally-enhanced’ B-scan.\\
\par Experimentally, the RMSE and perceptual loss when combined in a weighted-ratio of 1.0:0.01 offered the best performance (qualitative and quantitative; as described below).\\
\par The enhancer comprised of a total of 900 K trainable parameters, and was trained end-to-end using the Adam optimizer \cite{RN51}, with a learning rate of 0.0001. We trained and tested on an NVIDIA GTX 1080 founders edition GPU with CUDA 10.1 and cuDNN v7.5 acceleration. Using the given hardware configuration, the DL network enhanced a single ‘baseline’ B-scan in under 25 ms.\\

\subsection{Image Enhancement – Qualitative Analysis}

\indent Upon training, the network was used to enhance the unseen baseline B-scans from all the three devices. The DL-enhanced B-scans were qualitatively assessed by two expert observers (S.K.D and T.P.H) for the following: \textbf{(1)} noise reduction, \textbf{(2)} deep tissue visibility and blood vessel shadows, \textbf{(3)} contrast enhancement and intensity inhomogeneity, and \textbf{(4)} DL induced artifacts.\\

\subsection{Image Enhancement – Quantitative Analysis}
The following metrics were used to quantitatively assess the performance of the enhancer: \textbf{(1)} universal image quality index (UIQI) \cite{RN52}, and \textbf{(2)} structural similarity index (SSIM) \cite{RN53}. We used the UIQI to assess the extent of image enhancement (baseline vs. DL-enhanced B-scans), while the MSSIM was used to assess the structural reliability of the DL-enhanced B-scans (digitally-enhanced vs. DL-enhanced).\\
\par Unlike the traditional error summation methods (e.g., RMSE etc.) that compared only the intensity differences, the UIQI jointly modeled the \textbf{(1)} loss of correlation $(L_C)$, \textbf{(2)} luminance distortion $(D_L)$, and \textbf{(3)} contrast distortion $(D_C)$ to assess image quality \cite{RN52}. It was defined as (x: baseline; y: DL-enhanced B-scan):\\

\begin{align*}
UIQI(x,y)=L_C \times D_L \times D_C
\end{align*}

where, 

\begin{align*}
L_C = \frac{\sigma_{xy}}{\sigma_x \sigma_y }; 
D_L= \frac{2 \mu_{x}\mu_{y}}{\mu_{x}^2 + \mu_{y}^2 }; 
D_C= \frac{2 \sigma_{x}\sigma_{y}}{\sigma_{x}^2 + \sigma_{y}^2 }
\end{align*}

$(L_C)$ measured the degree of linear correlation between the baseline and DL-enhanced B-scans; $(D_L)$ and  $(D_C)$ measured the distortion in luminance and contrast respectively; $\mu_x$, $\sigma_x$, $\sigma_{x}^2$  denoted the mean, standard deviation, and variance of the intensity for B-scan x, while $\mu_{y}$, $\sigma_{y}$, $\sigma_{y}^2$  denoted the same for the B-scan y; $\sigma_{xy}$ was the cross-covariance between the two B-scans. The UIQI was defined between -1 (poor quality) and +1 (excellent quality).
As in our previous study \cite{RN37}, the SSIM (x: digitally-enhanced; y: DL-enhanced B-scan) was defined as: 

\begin{align*}
SSIM(x,y)=\frac{(2\mu_{x}\mu_{y}+C_1)(2\sigma_{xy}+C_2)}{(\mu_{x}^2+ \mu_{y}^2 + C_1)(\sigma_{x}^2+ \sigma_{y}^2 + C_2)}
\end{align*}

The constants $C_1$ and $C_2$  (to stabilize the division) were chosen as 6.50 and 58.52, as recommended in a previous study \cite{RN53}. The SSIM was defined between -1 (no similarity) and +1 (perfect similarity).

\subsection{3D Segmentation – Dataset Preparation}

The 60 volumes used for training and testing the 3D segmentation framework (20 from each device, balanced with respect to pathology) were manually segmented (slice-wise) by an expert observer (SD) using Amira (version 6, FEI, Hillsboro, OR). The following classes of tissues were segmented (\textbf{Figure \ref{fig:figure_1} B}): \textbf{(1)} the RNFL and prelamina (in red), \textbf{(2)} the ganglion cell complex (GCC; ganglion cell layer + inner plexiform layer; in cyan), \textbf{(3)} all other retinal layers (in blue); \textbf{(4)} the retinal pigment epithelium (RPE; in pink); \textbf{(5)} the choroid (in yellow); and \textbf{(6)} the lamina cribrosa (LC; in indigo). Noise (all regions below the choroid-sclera interface; in grey) and vitreous humor (black) were also isolated. We were unable to obtain a full-thickness segmentation of the LC due to limited visibility \cite{RN54}. We also excluded the peripapillary sclera due to its poor visibility and the extreme subjectivity of its boundaries especially in Cirrus and RTVue volumes. To optimize computational speed, the volumes (baseline OCT + labels) for all three devices were resized (in voxels) to 112 (height) x 88 (width) x 48 (number of B-scans).

\subsection{Deep Learning Based 3D Segmentation of the ONH }

Recent studies have demonstrated that 3D CNNs can improve the reliability of automated segmentation \cite{RN55,RN56,RN57,RN58,RN59,RN60,RN61,RN62}, and even out-perform their 2D variants \cite{RN56}. This is because 3D CNNs not only harness the information from each image, but also effectively combine it with the depth-wise spatial information from adjacent images. Despite its tremendous potential, the applications of 3D CNNs  in ophthalmology is still in its infancy \cite{RN63,RN64,RN65,RN66,RN67,RN68}, and has not yet been explored for the segmentation of the ONH tissues.\\
\par Further, there exist discrepancies in the delineation of ambiguous regions (e.g., choroid-sclera boundary, LC boundary) even among different well-trained DL model depending upon the type and complexity of architecture/feature extraction, learning method, etc., causing variability in the automated measurements. To address this, recent DL studies have explored ensemble learning \cite{RN29,RN69,RN70,RN71,RN72,RN73,RN74,RN75,RN76,RN77}, a meta-learning approach that synergizes (combine and fine-tune) \cite{RN74} the predictions from multiple networks, to offer a single prediction that is closest to the ground-truth. Specifically, ensemble learning has shown to better generalize and increase the robustness of segmentations in OCT \cite{RN29,RN70} and other medical imaging modalities \cite{RN71,RN72,RN73,RN76}.\\
\par In this study, we designed and validated ‘ONH-Net’, a 3D segmentation framework inspired by the popular 3D U-Net \cite{RN57} to isolate six ONH tissues from OCT volumes. The ONH-Net consisted of three segmentation networks (3D CNNs) and one 3D CNN for ensemble learning (referred to as the ‘ensembler’). Each of the three segmentation CNNs offered an equally plausible segmentation, which were then synergized by the ‘ensembler’ to yield the final 3D segmentation of the ONH tissues. 

\subsection{3D Segmentation – Network Description }

\begin{figure}[H]
    \centering
    \includegraphics[scale=0.25]{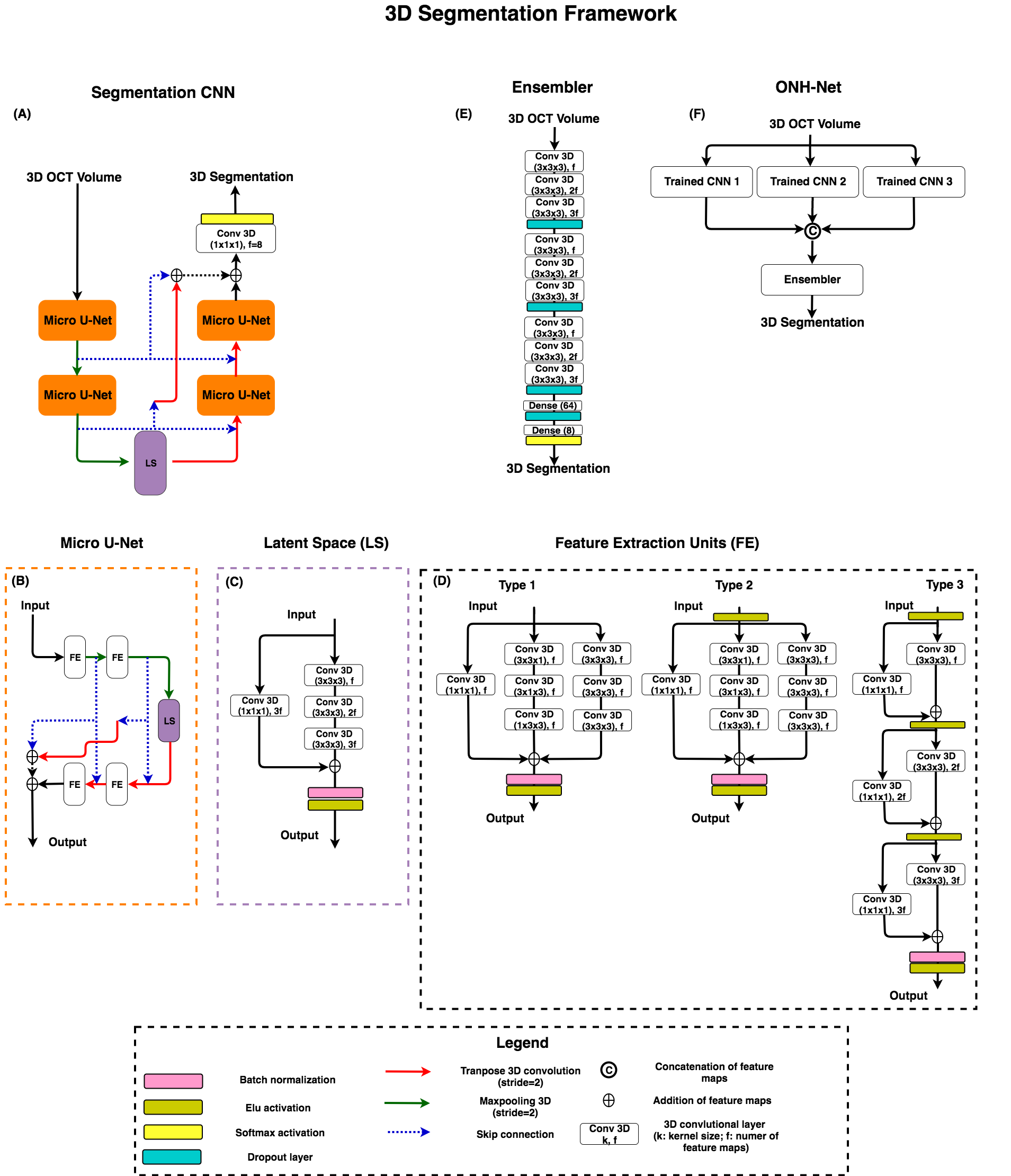}
    \caption{The DL architecture of the proposed 3D segmentation framework (three segmentation CNNs + one ensembler network) is shown. Each CNN \textbf{(A)} comprised of four micro-U-Nets ($\mu$-U-Nets; \textbf{(B)}) and a latent space (LS; \textbf{(C)}). The three CNNs differed from each other only in the design of the 'feature extraction' (FE) units (\textbf{(D); Types 1-3}). The ensembler \textbf{(E)} consisted of three sets of 3D convolutional layers, with each set separated by a dropout layer. ONH-Net \textbf{(F)} was then assembled by using the three trained CNNs as parallel input pipelines to the ensembler network. }
    \label{fig:figure_2}
\end{figure}

The design of the three segmentation CNNs was based on the 3D U-Net \cite{RN57} and its variants \cite{RN64}. Each of the three segmentation CNNs (\textbf{Figure \ref{fig:figure_2} A}) comprised of four micro-U-Nets ($\mu$-U-Nets; \textbf{Figure \ref{fig:figure_2} B}), and a latent space (\textbf{Figure \ref{fig:figure_2} C}).Both the segmentation CNNs and the $\mu$-U-Nets followed a similar design style.\\
 \par They consisted of an encoder segment that extracted contextual features (i.e. spatial arrangement of tissues), and a decoder segment that extracted the local information (i.e. tissue texture). The encoder segment sequentially downsampled the feature maps using the 3D max-pooling layers (stride=2,2,2), while the decoder segment sequentially upsampled using the 3D transposed convolutional layers (stride=2,2,2; filter size: 3x3x3; no of filters: 48).\\
 \par The latent space, implemented using residual blocks similar to our earlier study \cite{RN37}, transferred the extracted features from the encoder to the decoder segment. The use of residual learning improved the flow of gradient information through the network. 
 Skip connections \cite{RN38} between the encoder and decoder segments helped the DL network to jointly learn the contextual and local information, and the relationships between them.\\
 \par Also, as implemented in our earlier study \cite{RN37}, we used multi-scale hierarchical feature extraction to improve the delineation of tissue boundaries. The feature maps obtained from multi-scale hierarchical feature extraction were then added with the output of the decoder segment.\\
 \par The three segmentation CNNs deferred from each other only in the design of the 'feature extraction' (FE) units (\textbf{Figure \ref{fig:figure_2} D; Types 1-3}) that were used in the $\mu$-U-Nets.\\
 \par The Type 1 and Type 2 FE (\textbf{Figure \ref{fig:figure_2} D; Types 1-2}) units had a similar design, except that the input was pre-activated by the elu activation \cite{RN93} in Type 2 FE.\\
 \par In both the FE units, the input was passed through three parallel pathways: (1) the identity pathway; (2) the planar pathway; and (3) the volumetric pathway. The identity pathway implemented using a 1x1x1 3D convolutional layer allowed the unimpeded flow of gradient information throughout the network. In the planar pathway, the information from any two dimensions was extracted by the network at once (filter size: 3x3x1 [height x width]; 3x1x3 [height x depth]; 1x3x3 [width x depth]; 48 filters each). The volumetric pathway exploited the depth-wise spatially related and continuous information from all three dimensions at once (i.e., tissue morphology) using three 3D convolutional layers (filter size: 3x3x3; no of filters: 48).Finally, the feature maps from all the three pathways were added, batch normalized \cite{RN94}, and elu activated \cite{RN93}.\\
 \par In the Type 3 FE (\textbf{Figure \ref{fig:figure_2} D}) unit, the input was elu activated and passed on to three sets of simple residual blocks with 48, 96, and 144 filters, respectively. In each residual block, one 3D convolutional layer (filter size: 3x3x3) extracted the features, while a 1x1x1 3D convolution layer was used as the identity connection \cite{RN40}. The feature maps were then added, elu activated, and passed on to the next block. Finally, the feature maps were batch normalized and elu activated.\\
 \par For all three segmentation CNNs, the pre-final output feature maps (decoder output + multi-scale hierarchical feature extraction) were passed through a 3D convolutional layer (filter size: 1x1x1; no of filters: 8 [number of classes; 6 tissues + noise + vitreous humor]) and softmax activated to obtain the tissue-wise probability for each pixel. For each pixel, the tissue class of the highest probability was then assigned.\\
\par The ensembler (\textbf{Figure \ref{fig:figure_2} E}) was then implemented using three sets of 3D convolutional layers (specifications for each set; filter size [no of filters]: 3x3x3 [48]; 3x3x3 [96]; 3x3x3 [192]). A dropout \cite{RN78} of 0.50 was used between each set to reduce overfitting and improve the generalizability of the DL network. The feature maps were then passed through two dense layers of 64 and 8 units (number of classes) respectively, that were separated by a dropout layer (0.50). Finally, a softmax activation was applied to obtain the pixel-wise predictions.\\
\par Each of the three segmentation CNNs were first trained end-to-end with the same labeled-dataset. The ONH-Net was then assembled by using the three trained CNNs as parallel input pipelines to the ensembler network (\textbf{Figure \ref{fig:figure_2} F}). Finally, we trained the ONH-Net (ensembler weights: trainable; segmentation CNN weights: frozen) end-to-end using the same aforementioned labeled-dataset. During this process, each segmentation CNN provided equally plausible segmentation feature maps (obtained from the last 3D convolution layer), which were then concatenated and fed to the ensembler for fine-tuning. The ONH-Net was trained separately for each device. \\
\par All the DL networks (segmentation CNNs, ONH-Net) were trained with the stochastic gradient descent (SGD; learning rate:0.01; Nestrov momentum:0.05 \cite{RN79}) optimizer, and the Jaccard distance was used as the loss function \cite{RN24}. We empirically observed that the use of SGD optimizer with Nesterov momentum offered a better generalizability and faster convergence compared to Adam optimizer \cite{RN51} for OCT segmentation problems that typically use limited data, while Adam performed better for image-to-image translation problems (i.e., enhancement \cite{RN37}) that use much larger datasets. However, we are unable to theoretically explain this yet for our case. Given the limitations in hardware, all the DL networks were trained with a batch size of 1. To circumvent the scarcity in data, all the DL networks used custom data augmentation techniques (B-scans wise) as in our earlier study \cite{RN24}. We ensured that the same data augmentation was used for each B-scan in a given volume.\\
\par The three CNNs consisted of 7.2 M (Type 1), 7.2 M (Type 2), and 12.4 M (Type 3) trainable parameters, while the ONH-Net consisted of 28.86 M parameters (2.06M trainable parameters [ensembler], 26.8M non-trainable parameters [trained CNNs with weights frozen]). All the DL networks were trained and tested on NVIDIA GTX 1080 founders edition GPU with CUDA 10.1 and cuDNN v7.5 acceleration. Using the given hardware configuration, the ONH-Net was trained in 12 hours (10 hours for each CNN [trained in parallel; one per GPU]; 2 hours for fine-tuning with the ensembler). Once trained, each OCT volume was segmented in about 120 ms.\\

\subsection{3D Segmentation – Training and Testing}
We used a five-fold cross-validation approach (for each device) to train and test the performance of ONH-Net. In this process, the labeled-dataset (20 OCT volumes + manual segmentations) was split into five equal parts. One part (‘left-out’ set; 4 OCT volumes + manual segmentations) was used as the testing dataset, while the remaining four parts (16 OCT volumes + manual segmentations) were used as the training dataset.  The entire process was repeated five times, each with a different ‘left-out’ testing dataset (and corresponding training dataset). Totally, for each device, the segmentation performance was assessed on 20 OCT volumes (4 per validation; 5-fold cross-validation).

\subsection{3D Segmentation Performance – Qualitative Analysis}
The segmentations obtained from the trained ONH-Net on unseen data were manually reviewed by expert observers (S.D. and T.P.H) and compared against their corresponding manual segmentations.

\subsection{3D Segmentation Performance – Quantitative Analysis}
We used the following metrics to quantitatively assess the segmentation performance: \textbf{(1)} Dice coefficient (DC); \textbf{(2)} specificity (Sp); and \textbf{(3)} sensitivity (Sn). The metrics were computed in 3D for the following tissues: \textbf{(1)} the RNFL and prelamina; \textbf{(2)} the GCC; \textbf{(3)} all other retinal layers; \textbf{(4)} the RPE; and \textbf{(5)} the choroid. Given the subjectivity in the visibility of the posterior LC boundary \cite{RN54}, we excluded LC from quantitative assessment. Noise and vitreous humor were also excluded from quantitative assessment. The Dice coefficient (DC) was used to assess the spatial overlap between the manual and DL segmentations (between 0 [no overlap] and 1 [perfect overlap]). For each tissue, the DC was computed as:\\

\begin{align*}
\textrm{Dice score (DC)}= \frac{2 \times |D \cap M|}
{2 \times |D \cap M| + |D \setminus M| + |M \setminus D|}
\end{align*}

\par where D and M were the voxels that represented the chosen tissue in the DL segmented and the corresponding manually segmented volumes. Specificity (Sp) was used to assess the true negative rate of the segmentation framework, while sensitivity (Sn) was used to assess the true positive rate. They were defined as:\\

\begin{align*}
\textrm{Specificity (Sp)}= \frac{|\overline{D} \cap \overline{M}|}
{|\overline{M}|};
\textrm{  Sensitivity}= \frac{|{D} \cap {M}|}
{|{M}|}
\end{align*}

where $\overline{D}$ represented the voxels that did not belong to the chosen tissue in the DL segmented volume, while $\overline{M}$ represented the same in the corresponding manually segmented volume.

\subsection{3D Segmentation Performance – Effect of Image Enhancement}
To assess if image enhancement had an effect on segmentation performance, we trained and tested ONH-Net on the baseline and the DL-enhanced datasets. For both datasets, ONH-Net was trained on any one device (Spectralis/Cirrus/RTVue), but tested on all the three devices (Spectralis, Cirrus, and RTVue). Paired t-tests were used to compare the differences (means) in the segmentation performance (Dice coefficients, sensitivities, specificities; mean of all tissues) for both cases. 

\subsection{3D Segmentation Performance – Device Independency}
When tested on a given device (Spectralis/Cirrus/RTVue), paired t-tests were used to assess the differences (Spectralis vs. Cirrus; Cirrus vs. RTVue; RTVue vs. Spectralis) in the segmentation performance depending on the device used for training ONH-Net. The process was performed with both baseline and DL-enhanced datasets.

\subsection{3D Segmentation Clinical Reliability – Automated Parameter Extraction}
Upon obtaining the DL segmentations, two clinically relevant structural parameters that are crucial for the diagnosis of glaucoma: the (1) peripapillary RNFL thickness (p-RNFLT); and the (2) peripapillary GCC thickness (p-GCCT) were automatically extracted as in our earlier works.\\
\par For each volume in the testing dataset, a circular scan of diameter 3.4mm centered around the ONH \cite{RN95} was obtained. The p-RNFL thickness (global) was computed as the distance between the inner limiting membrane and the posterior RNFL boundary (mean of 360$^{\circ}$ measure). The p-GCT (global) was computed as the distance between the posterior RNFL boundary and the inner plexiform layer boundary (mean of 360$^{\circ}$ measure).\\
\par The intraclass correlation coefficients (ICCs) were obtained to compare the measurements computed from the DL and their corresponding manual segmentations for all cases. 
\pagebreak

\section{Results}

\subsection{Image Enhancement – Qualitative Analysis}

\begin{figure}[!htp]
    \centering
    \includegraphics[scale=0.13]{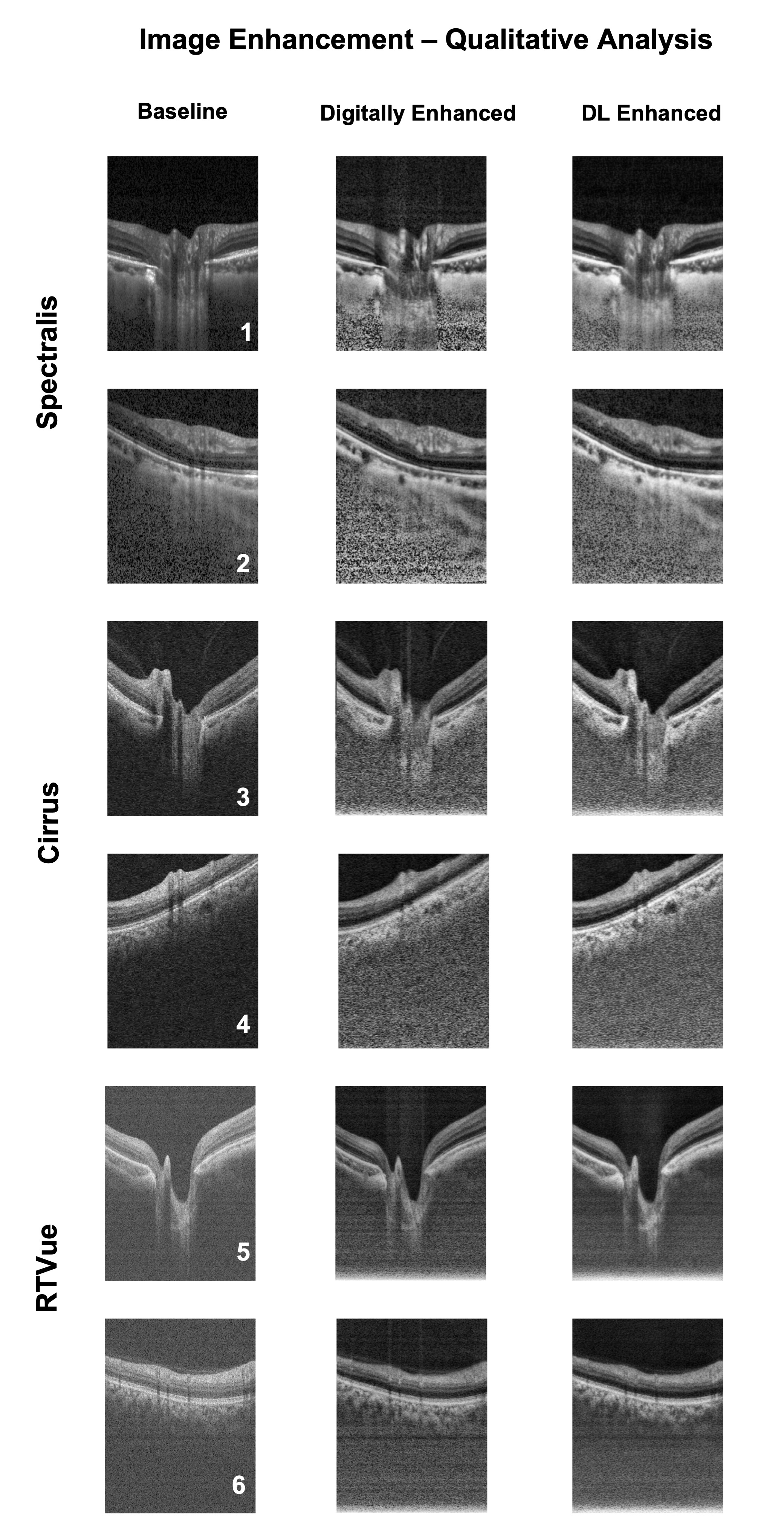}
    \caption{The qualitative performance of the image enhancement network is shown for six randomly selected \textbf{(1-6)} subjects (2 per device). The 1st, 2nd and 3rd columns represent the baseline, digitally-enhanced, and the corresponding DL-enhanced B-scans for patients imaged with Spectralis \textbf{(1-2)}, Cirrus \textbf{(3-4)}, and RTVue \textbf{(5-6)} devices, respectively.}
    \label{fig:figure_3}
\end{figure}

The enhancer was tested on a total of 1440 (480 from each device) unseen baseline B-scans. In the DL-enhanced B-scans from all the three devices (\textbf{Figure \ref{fig:figure_3}, Column 3}), the ONH-tissue boundaries appeared sharper with an uniformly enhanced intensity profile (compared to respective ‘baseline’ B-scans). The blood vessel shadows were also reduced with improved deep tissue (choroid-scleral interface, LC) visibility. In all cases, the DL-enhanced B-scans were consistently similar to their corresponding digitally-enhanced B-scans (\textbf{Figure \ref{fig:figure_3}, Column 2}), with no DL induced artifacts.   

\subsection{Image Enhancement – Quantitative Analysis}
The mean UIQI (mean $\pm$ SD) for the DL-enhanced B-scans (compared to baseline B-scans) were: 0.94 $\pm$ 0.02, 0.95 $\pm$ 0.03, and 0.97 $\pm$ 0.01 for Spectralis, Cirrus, and RTVue, respectively, indicating improved image quality.\\
\par In all cases, the mean SSIM (mean $\pm$ SD) for the DL-enhanced B-scans (compared to digitally-enhanced B-scans) were: 0.95 $\pm$ 0.02, 0.91 $\pm$ 0.02, and 0.93 $\pm$ 0.03, for Spectralis, Cirrus, and RTVue, respectively, indicating strong structural similarity.

\subsection{3D Segmentation Performance – Qualitative Analysis}
\begin{figure}[!htp]
    \centering
    \includegraphics[scale=0.130]{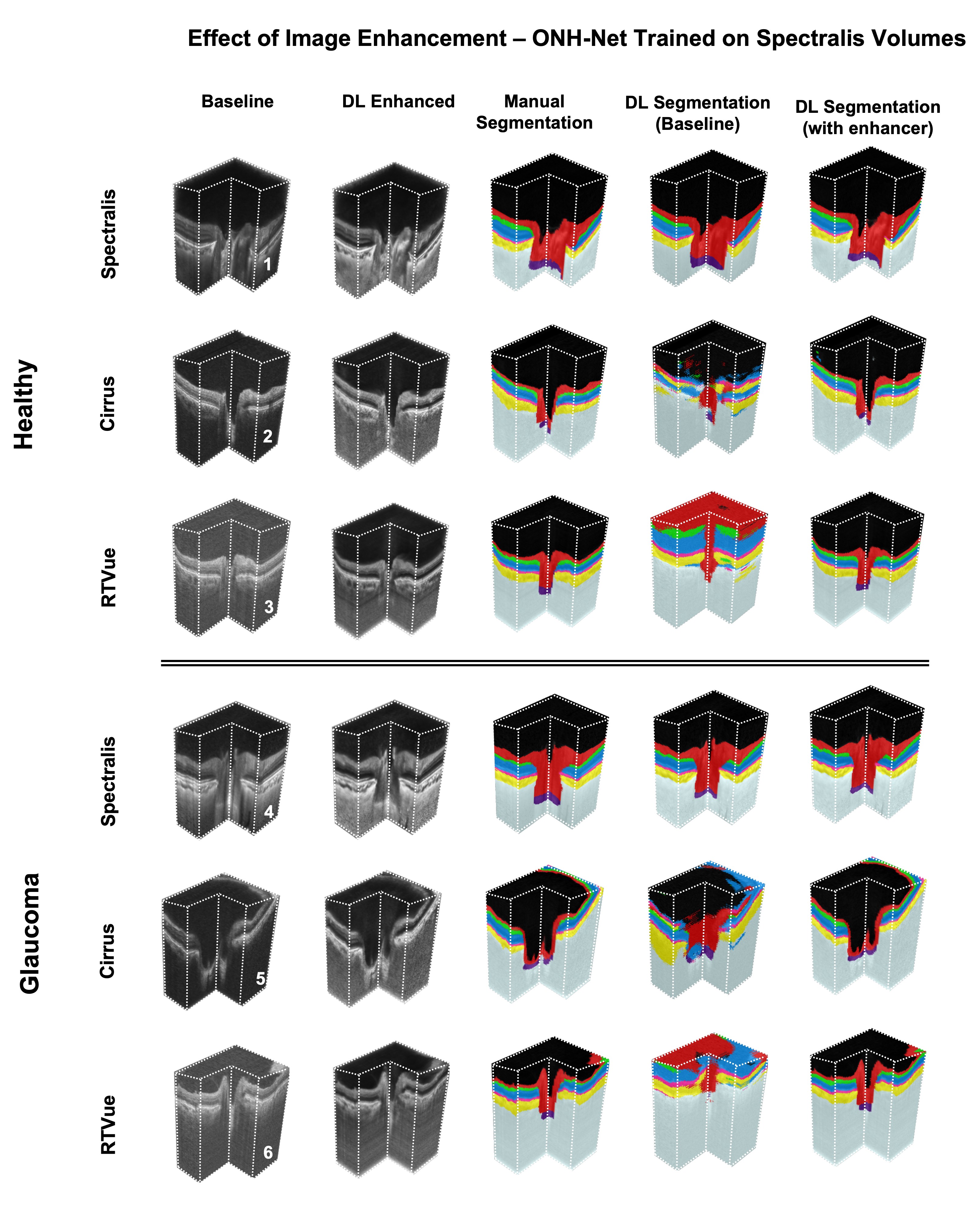}
    \caption{The segmentation performance (in 3D) on three healthy \textbf{(1-3)} and three glaucoma \textbf{(4-6)} subjects is shown. The ONH-Net was trained on volumes from Spectralis, and tested on Spectralis \textbf{(1, 4)}, Cirrus \textbf{(2, 5)}, and RTVue \textbf{(3,6)} devices respectively. The 1,st 2,nd and 3rd columns represent the baseline, DL enhanced, and the corresponding manual segmentations for the chosen volumes. The 4th and 5th columns represent the DL segmentations when the ONH-Net was trained with and without image enhancement respectively.}
    \label{fig:figure_10}
\end{figure}

When trained and tested on the baseline volumes from the same device \textbf{(\textbf{Figure \ref{fig:figure_4} - Figure \ref{fig:figure_6}; 4th column})}, ONH-Net successfully isolated all ONH layers. Further, the DL segmentations appeared consistent with their respective manual segmentations \textbf{(\textbf{Figure \ref{fig:figure_4} - Figure \ref{fig:figure_6}; 3rd column})}, with no difference in the segmentation performance between glaucoma and healthy OCT volumes. A representative case of the segmentations in 3D when trained on Spectralis and tested on the other three devices is shown in \textbf{Figure \ref{fig:figure_10}}. 

\subsection{3D Segmentation Performance – Quantitative Analysis}
When trained and tested on the baseline volumes (same device), the mean Dice coefficients (mean of all tissues; mean $\pm$ SD) were: 0.93 $\pm$ 0.02, 0.93 $\pm$ 0.02, and 0.93 $\pm$ 0.02 for Spectralis, Cirrus, and RTVue, respectively. The mean sensitivities / specificities (mean of all tissues; mean ± SD) were: 0.94 $\pm$ 0.02 / 0.99 $\pm$ 0.00, 0.93 $\pm$ 0.02 / 0.99 $\pm$ 0.00, and 0.93 $\pm$ 0.02 / 0.99 $\pm$ 0.00, respectively.

\subsection{3D Segmentation Performance – Effect of Image Enhancement and Device Independency}

\par
\begin{figure}[!htp]
    \centering
    \includegraphics[scale=0.32]{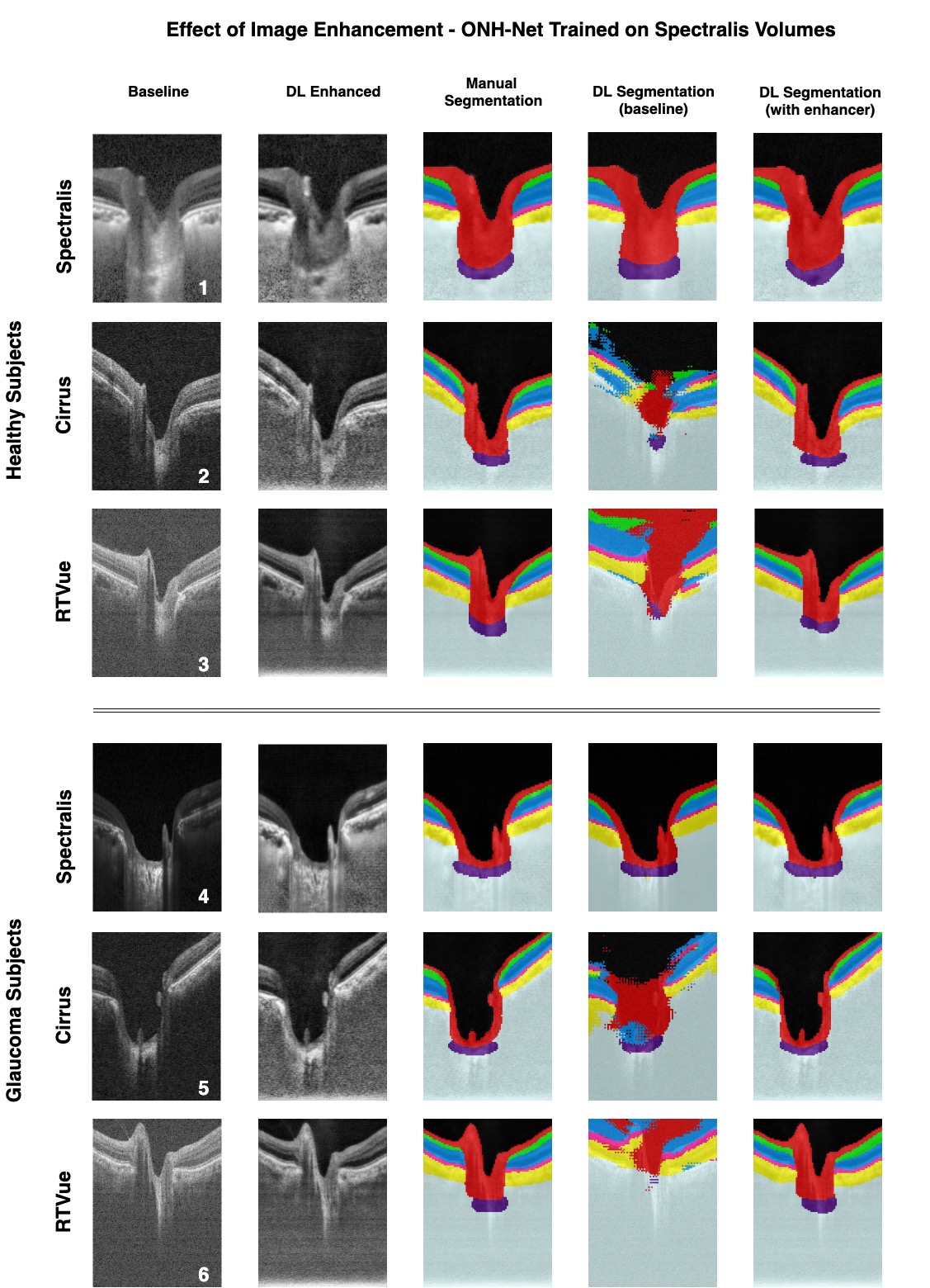}
    \caption{The qualitative performance (one randomly chosen B-scan per volume) of the ONH-Net 3D segmentation framework for three healthy \textbf{(1-3)} and three glaucoma \textbf{(4-6)} subjects is shown. The framework was trained on volumes from Spectralis, and tested on Spectralis \textbf{(1,4)}, Cirrus \textbf{(2,5)}, and RTVue \textbf{(3,6)} devices respectively. The 1st, 2nd and 3rd columns represent the baseline, DL enhanced, and the corresponding manual segmentation for the chosen B-scan. The 4th and 5th columns represent the DL segmentations when ONH-Net was trained and tested using the baseline and DL enhanced volumes, respectively. }
    \label{fig:figure_4}
\end{figure}

\begin{figure}[!htp]
    \centering
    \includegraphics[scale=0.32]{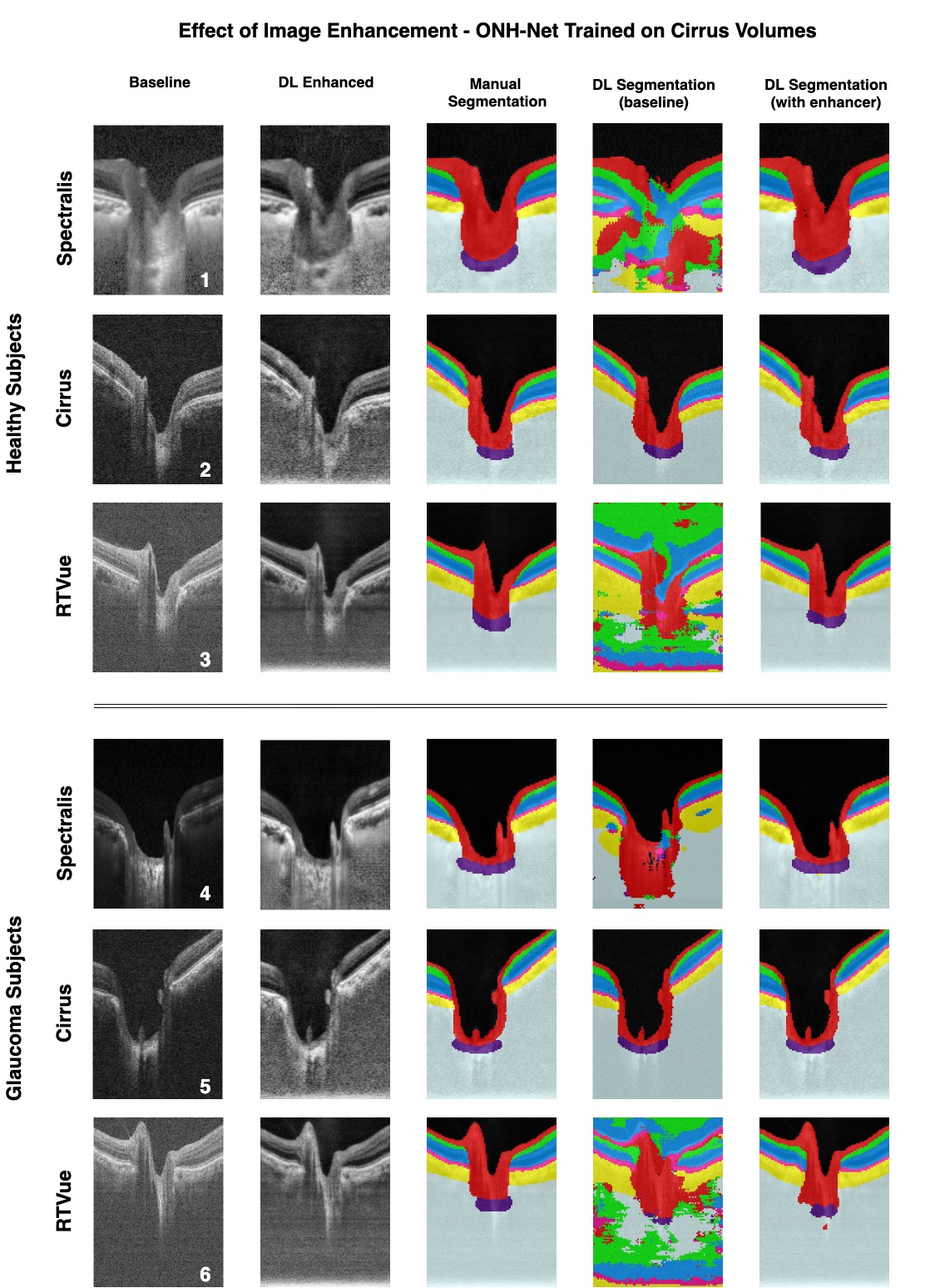}
    \caption{The qualitative performance (one randomly chosen B-scan per volume) of the ONH-Net 3D segmentation framework for three healthy \textbf{(1-3)} and three glaucoma \textbf{(4-6)} subjects is shown. The framework was trained on volumes from Cirrus, and tested on Spectralis \textbf{(1,4)}, Cirrus \textbf{(2,5)}, and RTVue \textbf{(3,6)} devices respectively. The 1st, 2nd and 3rd columns represent the baseline, DL enhanced, and the corresponding manual segmentation for the chosen B-scan. The 4th and 5th columns represent the DL segmentations when ONH-Net was trained and tested using the baseline and DL enhanced volumes, respectively. }
    \label{fig:figure_5}
\end{figure}

\begin{figure}[!htp]
    \centering
    \includegraphics[scale=0.32]{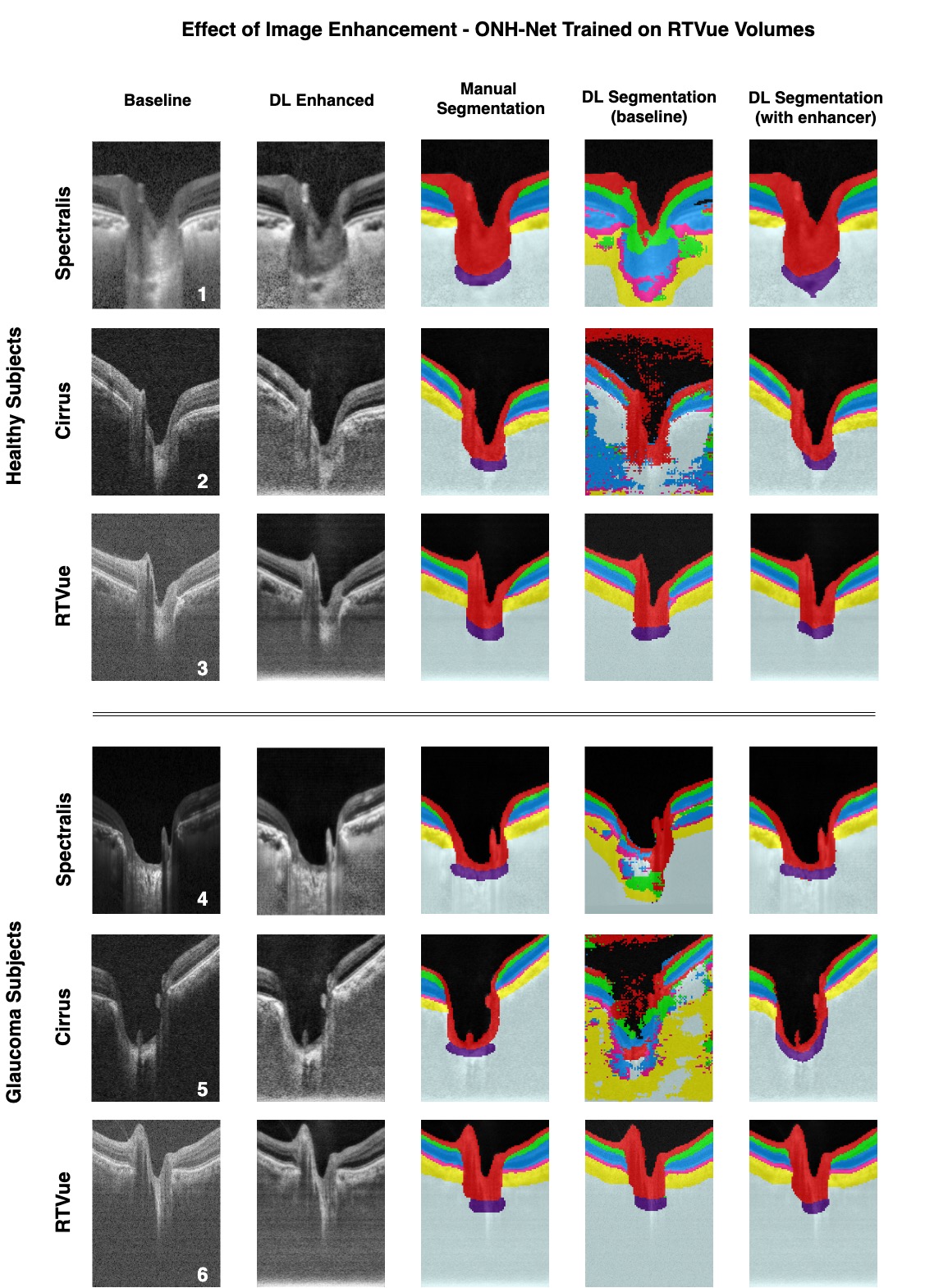}
    \caption{The qualitative performance (one randomly chosen B-scan per volume) of the ONH-Net 3D segmentation framework for three healthy \textbf{(1-3)} and three glaucoma \textbf{(4-6)} subjects is shown. The framework was trained on volumes from RTVue, and tested on Spectralis \textbf{(1,4)}, Cirrus \textbf{(2,5)}, and RTVue \textbf{(3,6)} devices respectively. The 1st, 2nd and 3rd columns represent the baseline, DL enhanced, and the corresponding manual segmentation for the chosen B-scan. The 4th and 5th columns represent the DL segmentations when ONH-Net was trained and tested using the baseline and DL enhanced volumes, respectively. }
    \label{fig:figure_6}
\end{figure}

 Without image enhancement (baseline dataset), ONH-Net trained with one device was unable to segment even a single ONH tissue reliably on the other two devices (\textbf{Figure \ref{fig:figure_4} ; Rows 2,3,5,6; Column 4}; similarly for \textbf{Figures \ref{fig:figure_5} - \ref{fig:figure_6}}). In all cases, dice coefficients were always lower than 0.65, sensitivities lower than 0.77, and specificities lower than 0.80.\\ 

\par However, with image enhancement (DL-enhanced dataset), ONH-Net trained with one device was able to accurately segment all tissue layers on the other two devices with mean Dice coefficients and sensitivities $>$ 0.92 (\textbf{Figures \ref{fig:figure_4}-\ref{fig:figure_6}, Column 5}). In addition, when trained and tested on the same device, it performed better for several ONH layers (p $<$ 0.05), when it was tested on the same device that it was trained on. The tissue wise quantitative metrics for the aforementioned cases can be found in \textbf{Tables \ref{tab:table_2}-\ref{tab:table_4}}.\\

\par
\begin{figure}[!htp]
    \centering
    \includegraphics[scale=0.33]{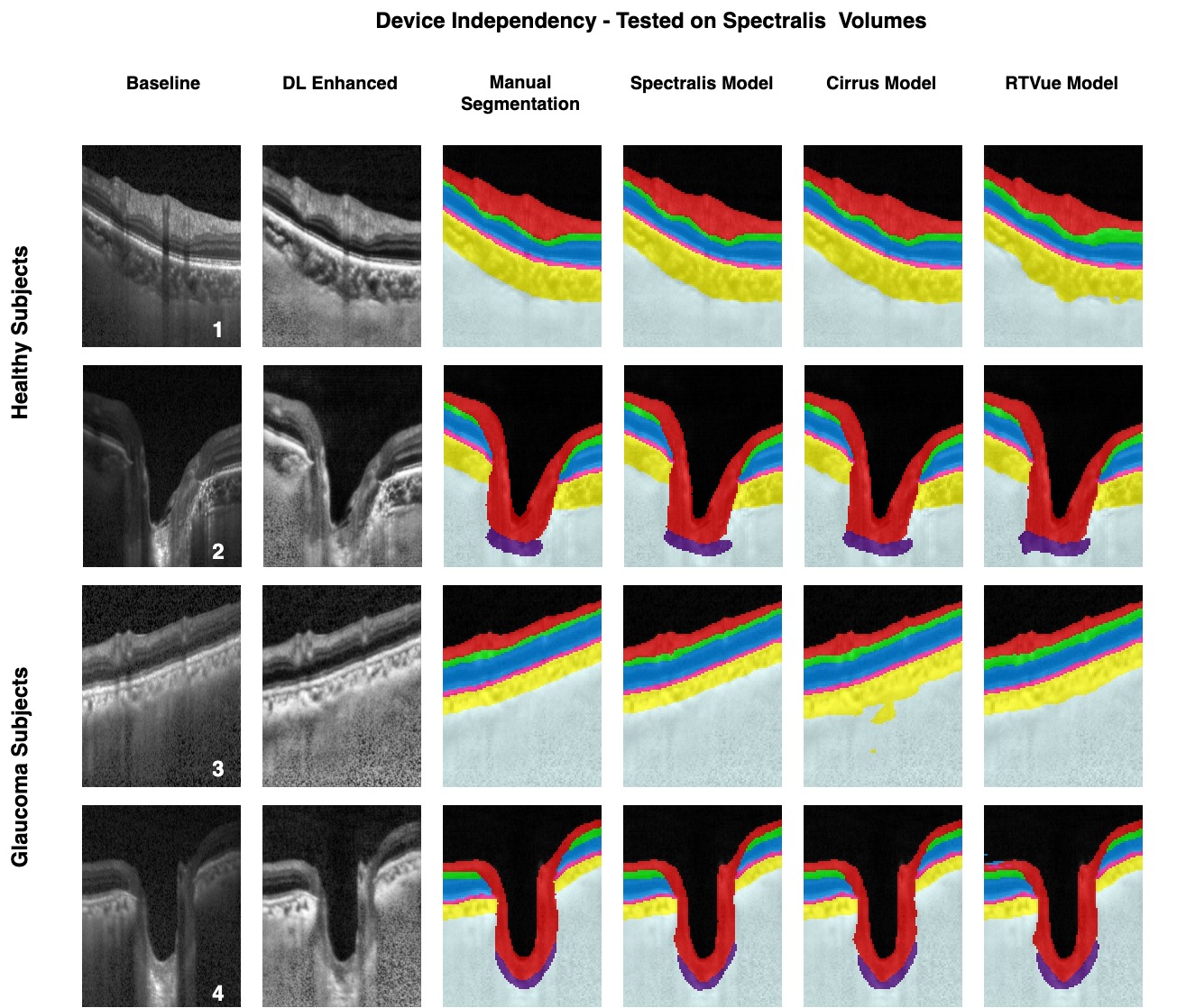}
    \caption{The device independent segmentation performance of the proposed ONH-Net is shown. The segmentation performance on four randomly chosen (\textbf{1-2} healthy; \textbf{3-4} glaucoma) Spectralis volumes from the test set are shown (one B-scan per volume). The 1st, 2nd, and 3rd columns represent the baseline, DL enhanced and the corresponding manual segmentation for the chosen B-scan. The 4th, 5th, and 6th columns represent the segmentations obtained when tested using the Spectralis, Cirrus, and RTVue trained segmentation model. }
    \label{fig:figure_7}
\end{figure}

\begin{figure}[!htp]
    \centering
    \includegraphics[scale=0.33]{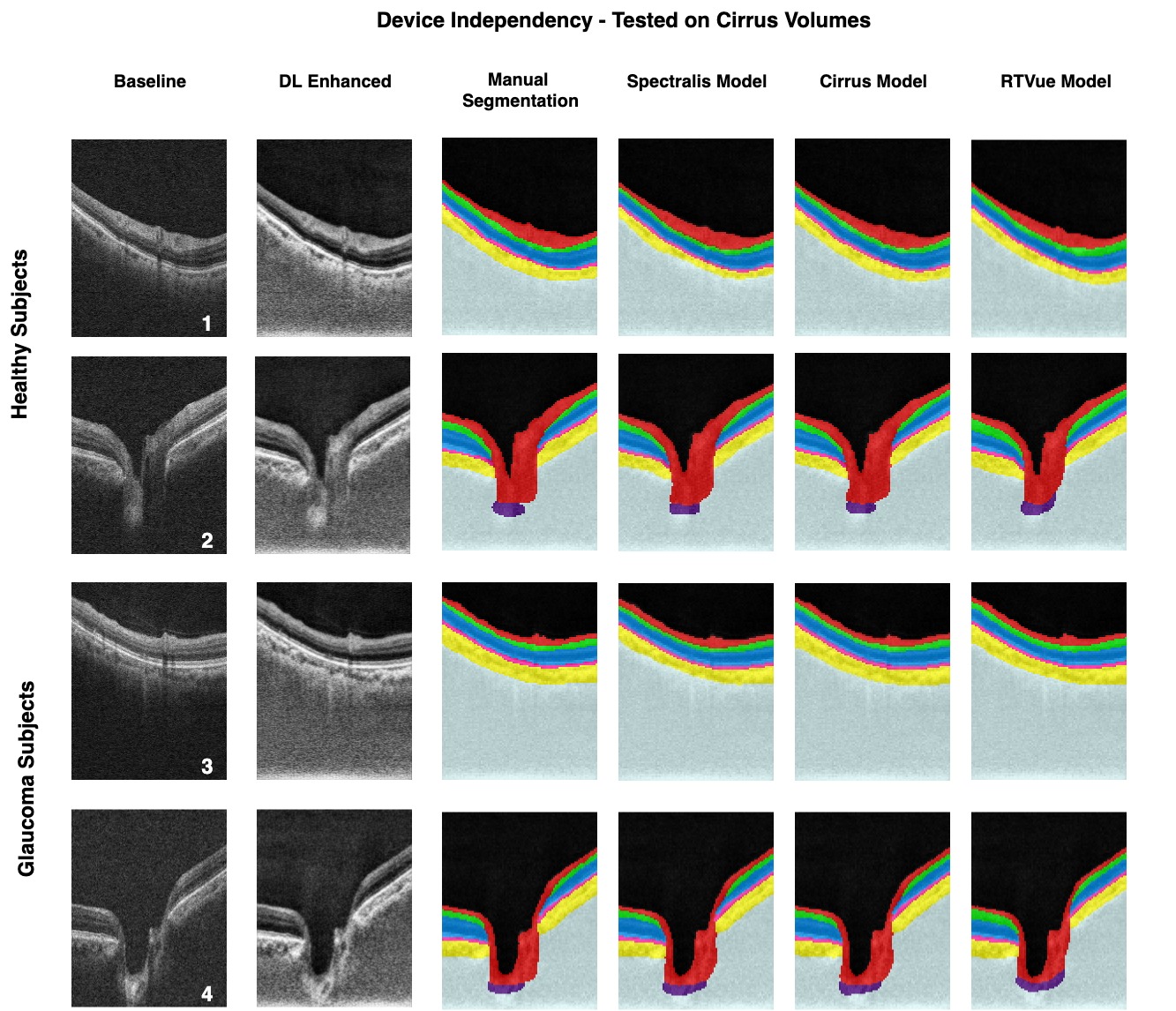}
    \caption{The device independent segmentation performance of the proposed ONH-Net is shown. The segmentation performance on four randomly chosen (\textbf{1-2} healthy; \textbf{3-4} glaucoma) Cirrus volumes from the test set are shown (one B-scan per volume). The 1st, 2nd, and 3rd columns represent the baseline, DL enhanced and the corresponding manual segmentation for the chosen B-scan. The 4th, 5th, and 6th columns represent the segmentations obtained when tested using the Spectralis, Cirrus, and RTVue trained segmentation model. }
    \label{fig:figure_8}
\end{figure}

\begin{figure}[!htp]
    \centering
    \includegraphics[scale=0.33]{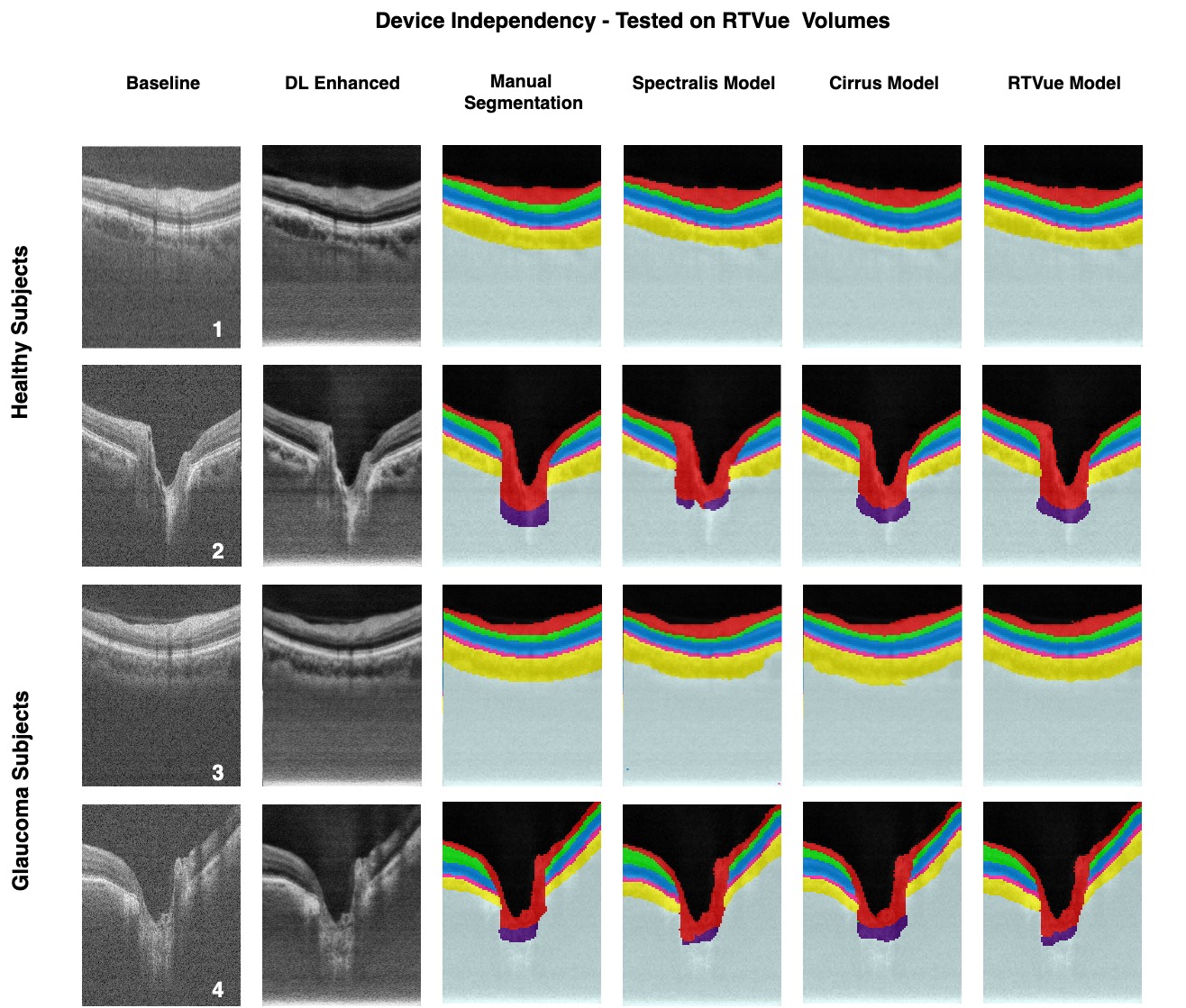}
    \caption{The device independent segmentation performance of the proposed ONH-Net is shown. The segmentation performance on four randomly chosen (\textbf{1-2} healthy; \textbf{3-4} glaucoma) RTVue volumes from the test set are shown (one B-scan per volume). The 1st, 2nd, and 3rd columns represent the baseline, DL enhanced and the corresponding manual segmentation for the chosen B-scan. The 4th, 5th, and 6th columns represent the segmentations obtained when tested using the Spectralis, Cirrus, and RTVue trained segmentation model. }
    \label{fig:figure_9}
\end{figure}

 Further, when trained and tested with the DL-enhanced OCT volumes, irrespective of the device used for training, there were no significant differences (p$<$0.05) in the segmentation performance for all tissues (\textbf{Figures \ref{fig:figure_7}-\ref{fig:figure_9}}), except for the LC.  The tissue wise quantitative metrics for the individual cases can be found in \textbf{Tables \ref{tab:table_5}-\ref{tab:table_7}}.\\

\subsection{3D Segmentation Clinical Reliability – Automated Parameter Extraction}
When trained and tested (same device) on the ‘baseline’ OCT volumes, the ICCs were always greater than 0.99 for both the p-RNFLT and the p-GCCT. However, when tested on the other two devices, since that the ONH-Net was unable to segment even a single tissue reliably, we did not extract the p-RNFLT and the g-GCCT for these cases.\\

\par When repeated the same with the ‘DL-enhanced’ volumes, irrespective of the device used for training, the ICCs were always greater than 0.98 for all cases, indicating excellent reliability.\\\\\\\\\\\\\\

\begin{table}[H]

\Large
\centering
\scalebox{0.48}{
\begin{tabular}{@{}cccccccccccc@{}}
\toprule
\multicolumn{12}{c}{\textbf{Effect of  Image Enhancement - Spectralis Trained Framework}} \\ \midrule
\multicolumn{2}{c}{\multirow{2}{*}{\textbf{Testing Device}}} &
  \multicolumn{2}{c}{\textbf{RNFL}} &
  \multicolumn{2}{c}{\textbf{GCC}} &
  \multicolumn{2}{c}{\textbf{Other Retinal Layers}} &
  \multicolumn{2}{c}{\textbf{RPE}} &
  \multicolumn{2}{c}{\textbf{Choroid}} \\ \cmidrule(l){3-12} 
\multicolumn{2}{c}{} &
  \textbf{w/o} &
  \textbf{w} &
  \textbf{w/o} &
  \textbf{w} &
  \textbf{w/o} &
  \textbf{w} &
  \textbf{w/o} &
  \textbf{w} &
  \textbf{w/o} &
  \textbf{w} \\ \midrule
\multirow{3}{*}{\textbf{Spectralis}} &
  \textbf{DC} &
  0.951±0.023 &
  0.954±0.017 &
  0.898±0.0216 &
  { \textbf{0.931±0.020}} &
  0.912±0.011 &
  { \textbf{0.936±0.010}} &
  0.896±0.010 &
  0.918±0.014 &
  0.902±0.044 &
  { \textbf{0.926± 0.031}} \\ \cmidrule(l){2-12} 
 &
  \textbf{Sn} &
  0.946±0.031 &
  { \textbf{0.960±0.026}} &
  0.900±0.0290 &
  { \textbf{0.946±0.019}} &
  0.917±0.009 &
  { \textbf{0.947±0.010}} &
  0.906±0.025 &
  { \textbf{0.938±0.022}} &
  0.915±0.054 &
  { \textbf{0.931± 0.038}} \\ \cmidrule(l){2-12} 
 &
  \textbf{Sp} &
  0.997±0.005 &
  0.993±0.001 &
  0.995±0.002 &
  0.996±0.003 &
  0.995±0.000 &
  0.995±0.001 &
  0.994±0.001 &
  0.994±0.002 &
  0.995±0.004 &
  0.996± 0.002 \\ \midrule
 &
   &
   &
   &
   &
   &
   &
   &
   &
   &
   &
   \\ \midrule
\multirow{3}{*}{\textbf{Cirrus}} &
  \textbf{DC} &
  0.511±0.029 &
  { \textbf{0.943±0.027}} &
  0.302±0.061 &
  { \textbf{0.919±0.032}} &
  0.587±0.035 &
  { \textbf{0.918±0.031}} &
  0.562±0.020 &
  { \textbf{0.918±0.019}} &
  0.477±0.055 &
  { \textbf{0.902± 0.033}} \\ \cmidrule(l){2-12} 
 &
  \textbf{Sn} &
  0.701±0.035 &
  { \textbf{0.955±0.024}} &
  0.560±0.0312 &
  { \textbf{0.899±0.024}} &
  0.737±0.022 &
  { \textbf{0.937±0.020}} &
  0.683±0.019 &
  { \textbf{0.920±0.023}} &
  0.687±0.057 &
  { \textbf{0.896± 0.043}} \\ \cmidrule(l){2-12} 
 &
  \textbf{Sp} &
  0.781±0.004 &
  { \textbf{0.988±0.000}} &
  0.702±0.005 &
  { \textbf{0.983±0.004}} &
  0.811±0.001 &
  { \textbf{0.992±0.004}} &
  0.797±0.002 &
  { \textbf{0.991±0.001}} &
  0.797±0.005 &
  { \textbf{0.991± 0.004}} \\ \midrule
 &
   &
   &
   &
   &
   &
   &
   &
   &
   &
   &
   \\ \midrule
\multirow{3}{*}{\textbf{RTVue}} &
  \textbf{DC} &
  0.323±0.048 &
  { \textbf{0.951±0.031}} &
  0.281±0.068 &
  { \textbf{0.898±0.030}} &
  0.288±0.060 &
  { \textbf{0.896±0.041}} &
  0.337±0.055 &
  { \textbf{0.912±0.029}} &
  0.357±0.059 &
  { \textbf{0.934± 0.028}} \\ \cmidrule(l){2-12} 
 &
  \textbf{Sn} &
  0.517±0.039 &
  { \textbf{0.936±0.032}} &
  0.465±0.063 &
  { \textbf{0.931±0.038}} &
  0.466±0.051 &
  { \textbf{0.913±0.038}} &
  0.525±0.043 &
  { \textbf{0.910±0.031}} &
  0.513±0.061 &
  { \textbf{0.901± 0.028}} \\ \cmidrule(l){2-12} 
 &
  \textbf{Sp} &
  0.678±0.006 &
  { \textbf{0.996±0.003}} &
  0.522±0.009 &
  { \textbf{0.992±0.006}} &
  0.598±0.005 &
  { \textbf{0.994±0.005}} &
  0.728±0.006 &
  { \textbf{0.994±0.003}} &
  0.710±0.010 &
  { \textbf{0.994± 0.005}} \\ \bottomrule
\end{tabular}%
}
\caption{The quantitative segmentation performance \textbf{(DC: Dice coefficient; Sn: sensitivity; Sp: specificity)} of ONH-Net with \textbf{(w)} and without \textbf{(w/o)} the use of image enhancement. ONH-Net was \textbf{trained on Spectralis}, and \textbf{tested on Spectralis, Cirrus, RTVue} devices. The metrics for each tissue that were significantly higher (p$<$0.05) when image enhancement was used are in bold.}
\label{tab:table_2}
\end{table}

\pagebreak

\hspace{10em}
\hspace{10em}
\hspace{10em}

\begin{table}[H]
\Huge
\centering
\scalebox{0.30}{
\begin{tabular}{cccccccccccc}
\hline
\multicolumn{12}{c}{\textbf{Effect of  Image Enhancement - Cirrus Trained Framework}} \\ \hline
\multicolumn{2}{c}{\multirow{2}{*}{\textbf{Testing Device}}} &
  \multicolumn{2}{c}{\textbf{RNFL}} &
  \multicolumn{2}{c}{\textbf{GCC}} &
  \multicolumn{2}{c}{\textbf{Other Retinal Layers}} &
  \multicolumn{2}{c}{\textbf{RPE}} &
  \multicolumn{2}{c}{\textbf{Choroid}} \\ \cline{3-12} 
\multicolumn{2}{c}{} &
  \textbf{w/o} &
  \textbf{w} &
  \textbf{w/o} &
  \textbf{w} &
  \textbf{w/o} &
  \textbf{w} &
  \textbf{w/o} &
  \textbf{w} &
  \textbf{w/o} &
  \textbf{w} \\ \hline
\multirow{3}{*}{\textbf{Spectralis}} &
  \textbf{DC} &
  0.744±0.033 &
  { \textbf{0.966±0.027}} &
  0.694±0.045 &
  { \textbf{0.935±0.020}} &
  0.699±0.039 &
  { \textbf{0.923±0.023}} &
  0.607±0.038 &
  { \textbf{0.932±0.018}} &
  0.542±0.041 &
  { \textbf{0.937± 0.029}} \\ \cline{2-12} 
 &
  \textbf{Sn} &
  0.741±0.037 &
  { \textbf{0.946±0.031}} &
  0.790±0.028 &
  { \textbf{0.934±0.021}} &
  0.814±0.033 &
  { \textbf{0.949±0.027}} &
  0.750±0.040 &
  { \textbf{0.932±0.032}} &
  0.773±0.048 &
  { \textbf{0.940± 0.034}} \\ \cline{2-12} 
 &
  \textbf{Sp} &
  0.785±0.004 &
  { \textbf{0.987±0.000}} &
  0.767±0.003 &
  { \textbf{0.992±0.002}} &
  0.858±0.003 &
  { \textbf{0.994±0.002}} &
  0.801±0.003 &
  { \textbf{0.995±0.003}} &
  0.847±0.005 &
  { \textbf{0.995± 0.002}} \\ \hline
 &
   &
   &
   &
   &
   &
   &
   &
   &
   &
   &
   \\ \hline
\multirow{3}{*}{\textbf{Cirrus}} &
  \textbf{DC} &
  0.912±0.023 &
  { \textbf{0.965±0.022}} &
  0.830±0.034 &
  { \textbf{0.920±0.027}} &
  0.899±0.029 &
  { \textbf{0.932±0.026}} &
  0.863±0.032 &
  { \textbf{0.913±0.020}} &
  0.867±0.039 &
  { \textbf{0.924± 0.025}} \\ \cline{2-12} 
 &
  \textbf{Sn} &
  0.936±0.031 &
  { \textbf{0.974±0.030}} &
  0.899±0.028 &
  0.918±0.026 &
  0.900±0.020 &
  { \textbf{0.922±0.021}} &
  0.907±0.026 &
  { \textbf{0.933±0.015}} &
  0.913±0.033 &
  0.923±  0.028 \\ \cline{2-12} 
 &
  \textbf{Sp} &
  0.971±0.002 &
  { \textbf{0.988±0.001}} &
  0.984±0.002 &
  { \textbf{0.994±0.002}} &
  0.991±0.004 &
  0.995±0.001 &
  0.991±0.002 &
  { \textbf{0.996±0.002}} &
  0.989±0.001 &
  0.993±  0.005 \\ \hline
 &
   &
   &
   &
   &
   &
   &
   &
   &
   &
   &
   \\ \hline
\multirow{3}{*}{\textbf{RTVue}} &
  \textbf{DC} &
  0.643±0.051 &
  { \textbf{0.936±0.035}} &
  0.541±0.055 &
  { \textbf{0.896±0.029}} &
  0.538±0.059 &
  { \textbf{0.917±0.028}} &
  0.645±0.060 &
  { \textbf{0.905±0.030}} &
  0.493±0.064 &
  { \textbf{0.910± 0.031}} \\ \cline{2-12} 
 &
  \textbf{Sn} &
  0.643±0.048 &
  { \textbf{0.925±0.033}} &
  0.642±0.059 &
  { \textbf{0.920±0.031}} &
  0.660±0.048 &
  { \textbf{0.928±0.022}} &
  0.725±0.051 &
  { \textbf{0.925±0.027}} &
  0.765±0.055 &
  { \textbf{0.926± 0.034}} \\ \cline{2-12} 
 &
  \textbf{Sp} &
  0.700±0.010 &
  { \textbf{0.980±0.003}} &
  0.668±0.010 &
  { \textbf{0.986±0.004}} &
  0.708±0.004 &
  { \textbf{0.988±0.003}} &
  0.781±0.005 &
  { \textbf{0.990±0.004}} &
  0.806±0.009 &
  { \textbf{0.990± 0.005}} \\ \hline
\end{tabular}%
}
\caption{The quantitative segmentation performance \textbf{(DC: Dice coefficient; Sn: sensitivity; Sp: specificity)} of ONH-Net with \textbf{(w)} and without \textbf{(w/o)} the use of image enhancement. ONH-Net was \textbf{trained on Cirrus}, and \textbf{tested on Spectralis, Cirrus, RTVue} devices. The metrics for each tissue that were significantly higher (p$<$0.05) when image enhancement was used are in bold.}
\label{tab:table_3}
\end{table}

\vspace*{5 cm}

\begin{table}[h]
\Huge
\centering
\scalebox{0.30}{
\begin{tabular}{cccccccccccc}
\hline
\multicolumn{12}{c}{\textbf{Effect of  Image Enhancement - RTVue Trained Framework}} \\ \hline
\multicolumn{2}{c}{\multirow{2}{*}{\textbf{Testing Device}}} &
  \multicolumn{2}{c}{\textbf{RNFL}} &
  \multicolumn{2}{c}{\textbf{GCC}} &
  \multicolumn{2}{c}{\textbf{Other Retinal Layers}} &
  \multicolumn{2}{c}{\textbf{RPE}} &
  \multicolumn{2}{c}{\textbf{Choroid}} \\ \cline{3-12} 
\multicolumn{2}{c}{} &
  \textbf{w/o} &
  \textbf{w} &
  \textbf{w/o} &
  \textbf{w} &
  \textbf{w/o} &
  \textbf{w} &
  \textbf{w/o} &
  \textbf{w} &
  \textbf{w/o} &
  \textbf{w} \\ \hline
\multirow{3}{*}{\textbf{Spectralis}} &
  \textbf{DC} &
  0.722±0.038 &
  { \textbf{0.966±0.029}} &
  0.656±0.041 &
  { \textbf{0.912±0.028}} &
  0.667±0.033 &
  { \textbf{0.918±0.026}} &
  0.620±0.039 &
  { \textbf{0.915±0.020}} &
  0.565±0.058 &
  { \textbf{0.915± 0.021}} \\ \cline{2-12} 
 &
  \textbf{Sn} &
  0.717±0.037 &
  { \textbf{0.951±0.035}} &
  0.682±0.039 &
  { \textbf{0.915±0.028}} &
  0.690±0.040 &
  { \textbf{0.931±0.031}} &
  0.620±0.038 &
  { \textbf{0.917±0.036}} &
  0.710±0.044 &
  { \textbf{0.922± 0.033}} \\ \cline{2-12} 
 &
  \textbf{Sp} &
  0.802±0.007 &
  { \textbf{0.986±0.003}} &
  0.707±0.005 &
  { \textbf{0.994±0.003}} &
  0.708±0.005 &
  { \textbf{0.996±0.001}} &
  0.706±0.004 &
  { \textbf{0.996±0.003}} &
  0.732±0.005 &
  { \textbf{0.985± 0.006}} \\ \hline
 &
   &
   &
   &
   &
   &
   &
   &
   &
   &
   &
   \\ \hline
\multirow{3}{*}{\textbf{Cirrus}} &
  \textbf{DC} &
  0.702±0.026 &
  { \textbf{0.942±0.024}} &
  0.682±0.040 &
  { \textbf{0.939±0.024}} &
  0.601±0.035 &
  { \textbf{0.907±0.031}} &
  0.595±0.045 &
  { \textbf{0.909±0.029}} &
  0.499±0.054 &
  { \textbf{0.911± 0.027}} \\ \cline{2-12} 
 &
  \textbf{Sn} &
  0.722±0.034 &
  { \textbf{0.954±0.033}} &
  0.700±0.038 &
  { \textbf{0.916±0.028}} &
  0.734±0.035 &
  { \textbf{0.922±0.030}} &
  0.765±0.041 &
  { \textbf{0.901±0.035}} &
  0.667±0.043 &
  { \textbf{0.905± 0.025}} \\ \cline{2-12} 
 &
  \textbf{Sp} &
  0.755±0.009 &
  { \textbf{0.995±0.002}} &
  0.763±0.006 &
  { \textbf{0.993±0.004}} &
  0.698±0.008 &
  { \textbf{0.994±0.002}} &
  0.799±0.007 &
  { \textbf{0.984±0.008}} &
  0.725±0.005 &
  { \textbf{0.993± 0.002}} \\ \hline
 &
   &
   &
   &
   &
   &
   &
   &
   &
   &
   &
   \\ \hline
\multirow{3}{*}{\textbf{RTVue}} &
  \textbf{DC} &
  0.942±0.032 &
  0.951±0.028 &
  0.896±0.029 &
  { \textbf{0.911±0.017}} &
  0.895±0.021 &
  { \textbf{0.925±0.008}} &
  0.881±0.035 &
  { \textbf{0.917±0.018}} &
  0.925±0.020 &
  0.931± 0.022 \\ \cline{2-12} 
 &
  \textbf{Sn} &
  0.932±0.020 &
  { \textbf{0.947±0.010}} &
  0.908±0.025 &
  0.928±0.019 &
  0.915±0.020 &
  { \textbf{0.931±0.019}} &
  0.899±0.026 &
  { \textbf{0.929±0.024}} &
  0.890±0.031 &
  { \textbf{0.924± 0.020}} \\ \cline{2-12} 
 &
  \textbf{Sp} &
  0.976±0.009 &
  { \textbf{0.997±0.001}} &
  0.983±0.005 &
  0.991±0.003 &
  0.985±0.001 &
  0.995±0.001 &
  0.990±0.004 &
  0.995±0.003 &
  0.989±0.002 &
  0.991±  0.006 \\ \hline
\end{tabular}%
}
\caption{The quantitative segmentation performance \textbf{(DC: Dice coefficient; Sn: sensitivity; Sp: specificity)} of ONH-Net with \textbf{(w)} and without \textbf{(w/o)} the use of image enhancement. ONH-Net was \textbf{trained on RTVue}, and \textbf{tested on Spectralis, Cirrus, RTVue} devices. The metrics for each tissue that were significantly higher (p$<$0.05) when image enhancement was used are in bold.}
\label{tab:table_4}
\end{table}

\begin{table}[H]
\Huge
\centering
\scalebox{0.35}{
\begin{tabular}{ccccc}
\hline
\multicolumn{2}{c}{\multirow{2}{*}{\textbf{Tissue (Spectralis Volumes)}}} & \multicolumn{3}{c}{\textbf{Training Device}} \\ \cline{3-5} 
\multicolumn{2}{c}{}                         & \textbf{Spectralis} & \textbf{Cirrus} & \textbf{RTVue} \\ \hline
\multirow{3}{*}{\textbf{RNFL}} & \textbf{DC} & 0.954±0.017         & 0.966±0.027     & 0.966±0.029    \\ \cline{2-5} 
                               & \textbf{Sn} & 0.960±0.026         & 0.946±0.031     & 0.951±0.035    \\ \cline{2-5} 
                               & \textbf{Sp} & 0.993±0.001         & 0.987±0.000     & 0.986±0.003    \\ \hline
                               &             &                     &                 &                \\ \hline
\multirow{3}{*}{\textbf{GCC}}  & \textbf{DC} & 0.931±0.020         & 0.935±0.020     & 0.912±0.028    \\ \cline{2-5} 
                               & \textbf{Sn} & 0.946±0.019         & 0.934±0.021     & 0.915±0.028    \\ \cline{2-5} 
                               & \textbf{Sp} & 0.996±0.003         & 0.992±0.002     & 0.994±0.003    \\ \hline
                               &             &                     &                 &                \\ \hline
\multirow{3}{*}{\textbf{Other Retinal Layers}}        & \textbf{DC}       & 0.936±0.010   & 0.923±0.023   & 0.918±0.026  \\ \cline{2-5} 
                               & \textbf{Sn} & 0.947±0.010         & 0.949±0.027     & 0.931±0.031    \\ \cline{2-5} 
                               & \textbf{Sp} & 0.995±0.001         & 0.994±0.002     & 0.996±0.001    \\ \hline
                               &             &                     &                 &                \\ \hline
\multirow{3}{*}{\textbf{RPE}}  & \textbf{DC} & 0.918±0.014         & 0.932±0.018     & 0.915±0.020    \\ \cline{2-5} 
                               & \textbf{Sn} & 0.938±0.022         & 0.932±0.032     & 0.917±0.036    \\ \cline{2-5} 
                               & \textbf{Sp} & 0.994±0.002         & 0.995±0.003     & 0.996±0.003    \\ \hline
                               &             &                     &                 &                \\ \hline
\multirow{3}{*}{\textbf{Choroid}}                     & \textbf{DC}       & 0.926± 0.031  & 0.937± 0.029  & 0.915± 0.021 \\ \cline{2-5} 
                               & \textbf{Sn} & 0.931± 0.038        & 0.940± 0.034    & 0.922± 0.033   \\ \cline{2-5} 
                               & \textbf{Sp} & 0.996± 0.002        & 0.995± 0.002    & 0.985± 0.006   \\ \hline
\end{tabular}%
}
\caption{The device independent segmentation performance \textbf{(DC: Dice coefficient; Sn: sensitivity; Sp: specificity)} of ONH-Net (using DL-enhanced dataset). The volumes from Spectralis device were tested on three segmentation models (Spectralis, Cirrus, and RTVue trained).}
\label{tab:table_5}
\end{table}

\vspace*{2 cm}

\begin{table}[H]
\Huge
\centering
\scalebox{0.35}{
\begin{tabular}{ccccc}
\hline
\multicolumn{2}{c}{\multirow{2}{*}{\textbf{Tissue (Cirrus Volumes)}}} & \multicolumn{3}{c}{\textbf{Training Device}} \\ \cline{3-5} 
\multicolumn{2}{c}{}                         & \textbf{Spectralis} & \textbf{Cirrus} & \textbf{RTVue} \\ \hline
\multirow{3}{*}{\textbf{RNFL}} & \textbf{DC} & 0.943±0.027         & 0.965±0.022     & 0.942±0.024    \\ \cline{2-5} 
                               & \textbf{Sn} & 0.955±0.024         & 0.974±0.030     & 0.954±0.033    \\ \cline{2-5} 
                               & \textbf{Sp} & 0.988±0.000         & 0.988±0.001     & 0.995±0.002    \\ \hline
                               &             &                     &                 &                \\ \hline
\multirow{3}{*}{\textbf{GCC}}  & \textbf{DC} & 0.919±0.032         & 0.920±0.027     & 0.939±0.024    \\ \cline{2-5} 
                               & \textbf{Sn} & 0.899±0.024         & 0.918±0.026     & 0.916±0.028    \\ \cline{2-5} 
                               & \textbf{Sp} & 0.983±0.004         & 0.994±0.002     & 0.993±0.004    \\ \hline
                               &             &                     &                 &                \\ \hline
\multirow{3}{*}{\textbf{Other Retinal Layers}}      & \textbf{DC}     & 0.918±0.031   & 0.932±0.026   & 0.907±0.031  \\ \cline{2-5} 
                               & \textbf{Sn} & 0.937±0.020         & 0.922±0.021     & 0.922±0.030    \\ \cline{2-5} 
                               & \textbf{Sp} & 0.992±0.004         & 0.995±0.001     & 0.994±0.002    \\ \hline
                               &             &                     &                 &                \\ \hline
\multirow{3}{*}{\textbf{RPE}}  & \textbf{DC} & 0.918±0.019         & 0.913±0.020     & 0.909±0.029    \\ \cline{2-5} 
                               & \textbf{Sn} & 0.920±0.023         & 0.933±0.015     & 0.901±0.035    \\ \cline{2-5} 
                               & \textbf{Sp} & 0.991±0.001         & 0.996±0.002     & 0.984±0.008    \\ \hline
                               &             &                     &                 &                \\ \hline
\multirow{3}{*}{\textbf{Choroid}}                   & \textbf{DC}     & 0.902± 0.033  & 0.924± 0.025  & 0.911± 0.027 \\ \cline{2-5} 
                               & \textbf{Sn} & 0.896± 0.043        & 0.923± 0.028    & 0.905± 0.025   \\ \cline{2-5} 
                               & \textbf{Sp} & 0.991± 0.004        & 0.993± 0.005    & 0.993± 0.002   \\ \hline
\end{tabular}%
}
\caption{The device independent segmentation performance \textbf{(DC: Dice coefficient; Sn: sensitivity; Sp: specificity)} of ONH-Net (using DL-enhanced dataset). The volumes from Cirrus device were tested on three segmentation models (Spectralis, Cirrus, and RTVue trained).}
\label{tab:table_6}
\end{table}

\vspace*{4 cm}

\begin{table}[H]
\Huge
\centering
\scalebox{0.35}{
\begin{tabular}{ccccc}
\hline
\multicolumn{2}{c}{\multirow{2}{*}{\textbf{Tissue (RTVue Volumes)}}} & \multicolumn{3}{c}{\textbf{Training Device}} \\ \cline{3-5} 
\multicolumn{2}{c}{}                         & \textbf{Spectralis} & \textbf{Cirrus} & \textbf{RTVue} \\ \hline
\multirow{3}{*}{\textbf{RNFL}} & \textbf{DC} & 0.951±0.031         & 0.936±0.035     & 0.951±0.028    \\ \cline{2-5} 
                               & \textbf{Sn} & 0.936±0.032         & 0.925±0.033     & 0.947±0.010    \\ \cline{2-5} 
                               & \textbf{Sp} & 0.996±0.003         & 0.980±0.003     & 0.997±0.001    \\ \hline
                               &             &                     &                 &                \\ \hline
\multirow{3}{*}{\textbf{GCC}}  & \textbf{DC} & 0.898±0.030         & 0.896±0.029     & 0.911±0.017    \\ \cline{2-5} 
                               & \textbf{Sn} & 0.931±0.038         & 0.920±0.031     & 0.928±0.019    \\ \cline{2-5} 
                               & \textbf{Sp} & 0.992±0.006         & 0.986±0.004     & 0.991±0.003    \\ \hline
                               &             &                     &                 &                \\ \hline
\multirow{3}{*}{\textbf{Other Retinal Layers}}     & \textbf{DC}     & 0.896±0.041   & 0.917±0.028   & 0.925±0.008  \\ \cline{2-5} 
                               & \textbf{Sn} & 0.913±0.038         & 0.928±0.022     & 0.931±0.019    \\ \cline{2-5} 
                               & \textbf{Sp} & 0.994±0.005         & 0.988±0.003     & 0.995±0.001    \\ \hline
                               &             &                     &                 &                \\ \hline
\multirow{3}{*}{\textbf{RPE}}  & \textbf{DC} & 0.912±0.029         & 0.905±0.030     & 0.917±0.018    \\ \cline{2-5} 
                               & \textbf{Sn} & 0.910±0.031         & 0.925±0.027     & 0.929±0.024    \\ \cline{2-5} 
                               & \textbf{Sp} & 0.994±0.003         & 0.990±0.004     & 0.995±0.003    \\ \hline
                               &             &                     &                 &                \\ \hline
\multirow{3}{*}{\textbf{Choroid}}                  & \textbf{DC}     & 0.934± 0.028  & 0.910± 0.031  & 0.931± 0.022 \\ \cline{2-5} 
                               & \textbf{Sn} & 0.901± 0.028        & 0.926± 0.034    & 0.924± 0.020   \\ \cline{2-5} 
                               & \textbf{Sp} & 0.994± 0.005        & 0.990± 0.005    & 0.991± 0.006   \\ \hline
\end{tabular}%
}
\caption{The device independent segmentation performance \textbf{(DC: Dice coefficient; Sn: sensitivity; Sp: specificity)} of ONH-Net (using DL-enhanced dataset). The volumes from RTVue device were tested on three segmentation models (Spectralis, Cirrus, and RTVue trained).}
\label{tab:table_7}
\end{table}

\newpage
\newpage

\section{Discussion}

In this study, we proposed a 3D segmentation framework (ONH-Net) that is easily translatable across OCT devices in a label-free manner (i.e. without the need to manually re-segment data for each device). Specifically, we developed 2 sets of DL networks. The first (referred to as the ‘enhancer’) was able to enhance OCT image quality from 3 OCT devices, and harmonized image-characteristics across these devices. The second performed 3D segmentation of 6 important ONH tissue layers. We found that the use of the ‘enhancer’ was critical for our segmentation network to achieve device independency. In other words, our 3D segmentation network trained on any of 3 devices successfully segmented ONH tissue layers from the other two devices with high performance.\\

\par Our work suggests that it is possible to automatically segment OCT volumes from a new OCT device without having to re-train ONH-Net with manual segmentations from that device. Besides existing commercial SD-OCT manufacturers, the democratization and emergence of OCT as the clinical gold-standard for in vivo ophthalmic examinations \cite{RN80} has encouraged the entry of several new manufacturers to the market as well. Further, owing to advancements in imaging technology, there has been a rise of the next generation devices: swept-source \cite{RN81}, polarization sensitive \cite{RN82}, and adaptive optics \cite{RN83} based OCTs. Given that preparing reliable manual segmentations (training data) for OCT-based DL algorithms requires months of training for a skilled technician, and that it would take more than 8 hours of manual work to accurately segment just a single 3D volume for just a limited number of tissue layers (here 6), it will soon become practically infeasible to perform manual segmentations for all OCT brands, device models, generations, and applications. Furthermore, only a few research groups have successfully managed to exploit DL to fully-isolate ocular structures from 3D OCT images \cite{RN23,RN24,RN25,RN26,RN27,RN30}, and only for a very limited number of devices. There is therefore a strong need for a single DL segmentation framework that can easily be translated across all existing and future OCT devices, thus eliminating the excruciating task of preparing training datasets manually. Our approach provides a high-performing solution to that problem.Eventually, we believe, this could open doors for multi-device glaucoma management.\\

\par In this study, we found that the use of enhancer was crucial for ONH-Net to achieve device independency, in other words, the ability to segment OCT volumes from devices it had not been trained with earlier. This can be attributed to the design of the proposed DL networks that allowed a perception of visual information through a host of low-level (e.g. tissue boundaries) and high-level abstract features (e.g. speckle pattern, intensity, and contrast profile). When image enhancement was used as a pre-processing step, the enhancer not only improved the quality of low-level features, but also reduced differences in high-level abstract features across OCT devices, thus ‘deceiving’ ONH-Net into perceiving volumes from all three devices similarly. This enabled ONH-Net trained on the DL-enhanced OCT volumes from one device to successfully isolate the ONH tissues from the other two devices with very high performance (mean Dice coefficients $>$ 0.92). Note that such a performance is superior to that of our previous 2D segmentation framework that also had the additional caveat that it only worked on a single device \cite{RN24}. In addition, irrespective of the device used for training, there were no significant differences (p $>$ 0.05) in segmentation performance. In all cases, our DL segmentations were deemed clinically reliable.\\

\par In a recent landmark study, De Fauw et al \cite{RN70} proposed the idea of using device-independent representations (segmentation maps) for the diagnosis of retinal pathologies from OCT images. However, the study was not truly device-independent, as, even though the diagnosis network was device-independent, the segmentation network was still trained with multiple devices. Similarly, our approach may not truly be considered as device-independent. While ONH-Net is device-independent, the enhancer (on which ONH-Net relies on) needs to be trained with data for all considered devices. But this is a still a very acceptable option, because the enhancer only requires un-labeled images (i.e. non-segmented; ~100 OCT volumes) for any new device that is being considered. After which, automated segmentation can still be performed without ever needing manual segmentation for that new device. Such a task would require a few minutes rather than several weeks/months needed for manual segmentations.\\

\par Finally, the proposed approach should not be confused with ‘transfer learning’ \cite{RN84}, a DL technique gaining momentum in medical imaging \cite{RN73,RN85,RN86,RN87,RN88}. In this technique, a DL network is first pre-trained on large-size datasets (e.g. ImageNet \cite{RN50}), and when subsequently fine-tuned on a smaller dataset for the task of interest (e.g. segmentation), it re-uses the pre-trained knowledge (high-level representations [e.g. edges, shapes]) to generalize better. In our approach, the generalization of ONH-Net was achieved using the enhanced images, and not the actual knowledge of the enhancer network, thus keeping the learning of both the networks mutually exclusive, yet necessary.\\

\par There are several limitations to this study that warrant further discussion. First, we used only 20 volumes in total to test the segmentation performance for each device. Second, the study was performed only using spectral-domain OCT devices, but not swept-source. Third, although the enhancer simultaneously addressed multiple issues affecting image quality, we were unable to quantify the effect of each. Also, we were unable to quantify the extent to which the ‘DL-enhanced’ B-scans were harmonized. Fourth, we observed slight differences in LC curvature and LC thickness when the LC was segmented using ONH-Net trained on different devices (\textbf{Figure \ref{fig:figure_7}, Figure \ref{fig:figure_8}, Figure \ref{fig:figure_9}; 2nd and 4th rows}). Given the significance of LC morphology in glaucoma \cite{RN89}, this subjectivity could affect glaucoma diagnosis. This has yet to be tested.This is yet to be tested. Further, in a few B-scans (\textbf{Figure \ref{fig:figure_7}, Figure \ref{fig:figure_8}, Figure \ref{fig:figure_9}; 6th column}), we observed that the GCC segmentations were thicker when the ONH-Net was trained on volumes from RTVue device. These variabilities might limit a truly multi-device glaucoma management. We are currently exploring the use of advanced DL concepts such as semi-supervised learning \cite{RN100} to address these issues that may have occurred as a result of limited training data. \\
\par Finally, although ONH-Net was invariant to volumes with glaucoma, it is unclear if the same will be true in the presence of other conditions such as cataract \cite{RN90}, peripapillary atrophy \cite{RN91}, and high-mypoia \cite{RN92} that commonly co-exist with glaucoma.\\

\par In conclusion, we demonstrate as a proof of concept that it is possible to develop DL segmentation tools that are easily translatable across OCT devices without ever needing additional manual segmentation data. To the best of our knowledge, our work is the first of its kind to propose a framework that could increase the clinical adoption of DL tools and eventually simplify glaucoma management. Finally, we hope the proposed framework can help patients for the longitudinal follow-up on multiple devices, and encourage multi-center glaucoma studies also.\\

\section{Funding}
Singapore Ministry of Education Academic Research Funds Tier 1 (R-397-000-294-114 [MJAG]); Singapore Ministry of Education Academic Research Funds Tier 2 (R-155-000-183-112 [AHT]; R-397-000-280-112, R-397-000-308-112 [MJAG]); National University of Singapore (NUS) Young Investigator Award Grant (NUSYIA FY16 P16,  R-155-000-180-133 [AHT]); National Medical Research Council (NMRC/OFIRG/0048/2017 [LS]).

\section*{Disclosures}
Dr.Micha\"el J. A. Girard and Dr.Alexandre H. Thi\'{e}ry are co-founders of Abyss Processing.

\newpage

\bibliographystyle{plain}
\bibliography{3d_seg}

\end{document}